\documentclass[aps,pra,superscriptaddress,footinbib,twocolumn]{revtex4-2}

\usepackage{dcolumn}
\usepackage{bm}

\usepackage[T1]{fontenc}
\usepackage{color}
\usepackage{graphicx}
\usepackage[american]{babel}
\usepackage{ulem}

\usepackage{amsmath,amsthm,amsfonts,amssymb,amscd}
\usepackage{indentfirst}
\usepackage{float}
\usepackage[dvipsnames]{xcolor}
\usepackage[colorlinks=true,citecolor=ForestGreen,linkcolor=Mulberry,urlcolor=BlueViolet,hyperindex]{hyperref}
\usepackage{mathtools}
\usepackage{bbold}
\usepackage{units}
\usepackage{soul}
\usepackage{ulem}

\newcommand{\ket}[1]{\vert#1\rangle}

\newcommand{\ketbra}[2]{\vert #1 \rangle \langle #2 \vert}
\newcommand{\id}{\mathbb{1}}
\newcommand{\tr}{\text{\normalfont Tr}}

\newcommand{\CASkey}{CAS Key Laboratory of Quantum Information, University of Science
and Technology of China, Hefei, 230026, China}
\newcommand{\CAScenter}{CAS Center For Excellence in Quantum Information and QuantumPhysics, Hefei, 230026, China}
\newcommand{\vcq}{University of Vienna, Faculty of Physics, Vienna Center for Quantum Science and Technology (VCQ) and Research platform TURIS, Boltzmanngasse 5, 1090 Vienna, Austria}
\newcommand{\iqoqi}{Institute for Quantum Optics and Quantum Information (IQOQI), Austrian Academy of Sciences, Boltzmanngasse 3, A-1090 Vienna, Austria}
\newcommand{\HFlab}{Hefei National Laboratory, University of Science and Technology of China, Hefei 230088, China}
\newcommand{\cdl}{Christian Doppler Laboratory for Photonic Quantum Computer, 8 Faculty of Physics, University of Vienna, 1090 Vienna, Austria}
\newcommand{\unige}{D\'epartement de Physique Appliqu\'ee, Universit\'e de Gen\`eve, 1211 Gen\`eve, Switzerland}

\begin{document}

\title{Semi-device-independent certification of indefinite causal order \\ in a photonic quantum switch}

\author{Huan Cao}
\thanks{These two authors contributed equally to this work.}
\affiliation{\CASkey}
\affiliation{\CAScenter}
\affiliation{\vcq}

\author{Jessica Bavaresco}
\thanks{These two authors contributed equally to this work.}
\affiliation{\iqoqi}
\affiliation{\unige}

\author{Ning-Ning Wang}
\affiliation{\CASkey}
\affiliation{\CAScenter}
\affiliation{\HFlab}

\author{Lee A. Rozema}
\affiliation{\vcq}

\author{Chao Zhang}
\email{drzhang.chao@ustc.edu.cn}
\affiliation{\CASkey}
\affiliation{\CAScenter}
\affiliation{\HFlab}

\author{Yun-Feng Huang}
\email{hyf@ustc.edu.cn}
\affiliation{\CASkey}
\affiliation{\CAScenter}
\affiliation{\HFlab}

\author{Bi-Heng Liu}
\affiliation{\CASkey}
\affiliation{\CAScenter}
\affiliation{\HFlab}

\author{Chuan-Feng Li}
\email{cfli@ustc.edu.cn}
\affiliation{\CASkey}
\affiliation{\CAScenter}
\affiliation{\HFlab}

\author{Guang-Can Guo}
\affiliation{\CASkey}
\affiliation{\CAScenter}
\affiliation{\HFlab}

\author{Philip Walther}
\affiliation{\vcq}
\affiliation{\cdl}


\begin{abstract}
We report an experimental certification of indefinite causal order that relies only on the characterization of the operations of a single party. We do so in the semi-device-independent scenario with the fewest possible assumptions of characterization of the parties' local operations in which indefinite causal order can be demonstrated with the quantum switch. To achieve this result, we introduce the concept of semi-device-independent causal inequalities and show that the correlations generated in a photonic quantum switch, in which all parties are able to collect local outcome statistics, achieve a violation of this inequality of $224$ standard deviations. This result consists of the experimental demonstration of indefinite causal order with the fewest device-characterization assumptions to date.
\end{abstract}

\maketitle


\section{Introduction}

Quantum mechanics challenges the viewpoint that physical quantities are locally pre-defined independently of measurement~\cite{bell64}. In recent years, pioneering work in quantum information has shown that, by assuming local operations respect quantum mechanics but dropping the assumption that the events must occur in a definite order, we may also challenge the viewpoint of well-defined causality~\cite{chiribella13switch,oreshkov12}.While extending the quantum circuit formalism~\cite{nielsenchuang,chiribella08architecture} and the notion of quantum combs~\cite{kretschmann05,gustoki06,chiribella09,pollock18}, these works predicted processes with an indefinite causal order (ICO) that nevertheless do not lead to contradictions or paradoxes in the collected statistical data~\cite{oreshkov12}.

The study of such causal structures shows that ICO brings advantages to the performance of several quantum tasks~\cite{chiribella12,araujo14,feix15,guerin16,ebler18,quintino18,zhao20,bavaresco21a,bavaresco21b,renner21}. From a fundamental perspective, the investigation of quantum causal structures not only renews our understanding of causality in nature, but also helps to address the long-standing problem of reconciling quantum theory and general relativity in a theory of quantum gravity~\cite{hardy2007towards,christodoulou2019possibility,marletto2017gravitationally}. A well-studied process with ICO is the quantum switch~\cite{chiribella13switch}, upon which all experimental investigations of ICO to date have been based~\cite{procopio15,rubino17,goswami18,rubino22,goswami20,wei19,guo20,rubino21,taddei21}. 
There has been some discussion regarding what is to be considered a valid implementation of the quantum switch as proposed in Ref. \cite{chiribella13switch}, with some of the opinion that current experiments are simulations \cite{paunkovic2020causal,vilasini2022embedding,ormrod2022causal}, while others conclude that the experiments have an indefinite causal order \cite{oreshkov2019time} or at least have a quantifiable resource advantage \cite{fellous2022comparing}.
To avoid ambiguity, here we call these experimental implementations 'photonic' quantum switches.

An experimental certification of ICO that depends exclusively on the collected statistical data, and critically does not rely on any assumptions about the description of the local operations or the process, is called a device-independent certification. It can be achieved via the violation of a causal inequality~\cite{oreshkov12,branciard16}, similar to how entanglement can be device-independently certified through the violation of a Bell inequality~\cite{brunner14}. However, not all processes with ICO are able to generate noncausal correlations that can be observed in a device-independent way~\cite{araujo15,feix15}, one such example is the quantum switch ~\cite{araujo15}. Although there exist theoretical processes that are able to violate causal inequalities~\cite{oreshkov12,branciard16}, currently, no experimental implementations of them are known. Indefinite causal order has, on the other hand, been certified in a device-dependent scenario {in a photonic quantum switch}, where the operations of all parties must be fully characterized. In this scenario, certification can be achieved through a causal witness~\cite{araujo15}, analogous to an entanglement witness~\cite{terhal00}. 
All but one experimental demonstration of ICO to date~\cite{procopio15,rubino17,goswami18,goswami20,wei19,guo20,rubino21,taddei21} have critically relied on fully device-dependent assumptions, essentially assuming a perfect implementation of all local operations. One recent experiment was reported~\cite{rubino22} in which the measurements performed by the final party of the {photonic} quantum switch were treated device-independently, but still assuming a full characterization of the operations of the other two parties inside the switch.
It also employed device-dependent assumptions in the analysis of the initial target system, leaving as an open question whether a certification of ICO that relies on fewer assumptions would be possible. Recently, new theoretical proposals have positively answered this question ~\cite{bavaresco19,dourdent21}. 

Here, we experimentally confirm this stronger form of certification by only making assumptions about the characterization of the operations of \textit{a single party}---in a semi-device-independent scenario.
Our certification relies on strictly fewer device-characterization assumptions than previous implementations. Moreover, the assumptions upon which we rely are the minimal set of complete device-characterization assumptions in which the quantum switch demonstrates noncausal properties~\cite{bavaresco19}.
Hence, our semi-device-independent certification is optimal in this sense.
By extending the framework of Ref.~\cite{bavaresco19}, we introduce the concept of tailored semi-device-independent causal inequalities, whose violation certifies ICO, parallel to how the violation of a steering inequality~\cite{cavalcanti09,skrzypczyk14} certifies entanglement in a semi-device-independent way. We experimentally test our inequality by implementing a photonic quantum switch. We develop a compact interferometer array, which incorporates multiple-outcome instruments for all parties acting on the quantum switch. This enables each party to generate local outcomes. This novel design allows us to experimentally test our inequality, yielding a violation by more than $224$ standard deviations.


\section{Results}

\subsection{Semi-device-independent causal inequalities}

\begin{figure}
\begin{center}
    \includegraphics[width=\columnwidth]{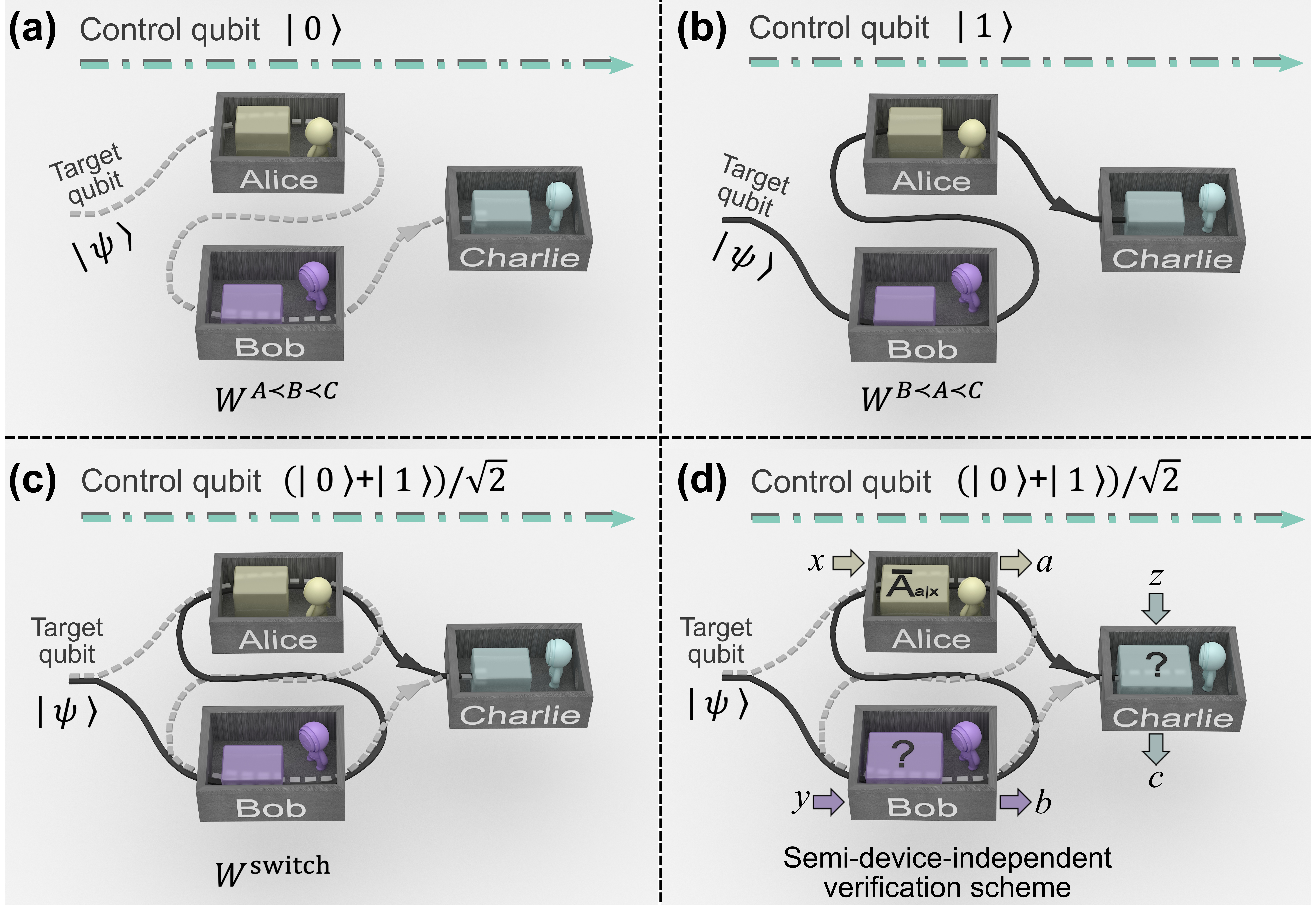}
    \caption{\textbf{Schematic diagram of the quantum switch.} (a)  and (b) Alice and Bob act on the target qubit in an order determined by the state of the control qubit, either $\ket{0}$ or $\ket{1}$. (c) When the control qubit is in the superposition state $\frac{1}{\sqrt{2}}(\ket{0}+\ket{1})$, Alice and Bob act on the target qubit in an ICO. (d) We certify the indefinite causal properties of the quantum switch in a semi-device-independent scenario, where only the operations of Alice are characterized while no assumptions are made about the operations of Bob and Charlie.}
    \label{fig:diagram}
\end{center}    
\end{figure}

The quantum switch~\cite{chiribella13switch} is a process that describes the following experimental situation. Consider an experiment in which two local parties, Alice and Bob, act on a qubit target system in an order determined by the state of a qubit control system (see Fig.~\ref{fig:diagram}). If the control system is in the state $\ket{0}$ ($\ket{1}$) Alice will act on the target system before (after) Bob. However, if the control system is in the coherent superposition $\ket{+}\coloneqq(\ket{0}+\ket{1})/\sqrt{2}$, then the target state will be acted upon by Alice and Bob in an \textit{indefinite causal order}. Finally, a third party, Charlie, that is in the well-defined future of Alice and Bob, performs a measurement in both target and control systems, regardless of the causal order between Alice and Charlie. Such a process, depicted in Fig.~\ref{fig:diagram}(d), allows for the events marked by the local operations of Alice and Bob to occur in what can be interpreted as a superposition of causal orders.

In the process matrix formalism~\cite{oreshkov12}, the quantum switch can be expressed as an operator~\cite{araujo15} $W^\text{switch}\in\mathcal{L}(\mathcal{H}^{A_I}\otimes\mathcal{H}^{A_O}\otimes\mathcal{H}^{B_I}\otimes\mathcal{H}^{B_O}\otimes\mathcal{H}^{C_t}\otimes\mathcal{H}^{C_c})$ that acts on the linear spaces of Alice and Bob's input ($A_I,B_I$) and output ($A_O,B_O$) systems, and on Charlie's input, where he receives the future states of the target ($C_t$) and control ($C_c$) systems. 
The fact that the quantum switch exhibits an ICO is formalized by the statement that this process cannot be expressed as a classical mixture of a process $W^{A\prec B\prec C}$  (Alice acting before Bob, and Bob before Charlie), with a process $W^{B\prec A\prec C}$ (Bob acting before Alice, and Alice before Charlie)~\cite{araujo15}. That is,
\begin{equation}\label{eq::nonsep}
    W^\text{switch} \neq q W^{A\prec B\prec C} + (1-q) W^{B\prec A\prec C},
\end{equation}
for any $0\leq q\leq 1$. This property is called \textit{causal nonseparability}~\cite{araujo15}, and is also referred to simply as indefinite causal order. 

Causal nonseparability can be certified through the correlations that arise when the independent parties involved in the experiment collect local statistics by making different choices of operations and recording their outcomes. Such operations are modelled by quantum instruments, which are the most general operations that can measure and transform a quantum system. Then, the joint probability distributions over the outcomes of Alice, Bob, and Charlie are given by a function of their instrument elements and the quantum switch process, according to a generalized Born rule~\cite{oreshkov12}
\begin{equation}\label{eq::bornrule}
    p(abc|xyz) = \tr\left[\left(A_{a|x}\otimes B_{b|y}\otimes M_{c|z}\right)\,W^\text{switch}\right],
\end{equation}
where $\{x,y,z\}$ label the inputs, $\{a,b,c\}$ label the outputs, and $A_{a|x}\in\mathcal{L}(\mathcal{H}^{A_I}\otimes\mathcal{H}^{A_O})$, $B_{b|y}\in\mathcal{L}(\mathcal{H}^{B_I}\otimes\mathcal{H}^{B_O})$, and $M_{c|z}\in\mathcal{L}(\mathcal{H}^{C_t}\otimes\mathcal{H}^{C_c})$ are the instrument elements of Alice and Bob, and the measurements of Charlie, respectively. Here, instruments are represented in the Choi picture~\cite{depillis67,jamiolkowski72,choi75}.

Following the framework developed in Ref.~\cite{bavaresco19}, consider an experiment that generates the correlations $\{p(abc|xyz)\}$ in a semi-device-independent scenario, by having Alice, Bob, and Charlie act on an uncharacterized process with a set of known operations $\{\bar{A}_{a|x}\}$ for Alice, and unknown operations for Bob and Charlie. 
This experiment certifies ICO if and only if, for some $a,b,c,x,y,z$,
\begin{equation}
    p(abc|xyz) \neq \tr[(\bar{A}_{a|x}\otimes B_{b|z}\otimes M_{c|z})W^\text{SEP}],
\end{equation}
for all sets of quantum instruments $\{B_{b|z}\}$ and measurements $\{M_{c|z}\}$, and for all tripartite causally separable process matrices $W^\text{SEP}$ that have a well-defined last party (i.e. of the form of the r.h.s. of Eq.~\eqref{eq::nonsep}). That is, ICO is certified when the experimentally measured correlations cannot be explained by a causally separable process regardless of the operations of Bob and Charlie, assuming only knowledge of Alice's operations. This statement is equivalent to showing that the experiment described by $\{p(abc|xyz)\}$ and $\{\bar{A}_{a|x}\}$ cannot be explained by a semi-device-independent causal model or causal assemblage~\cite{bavaresco19}.
A causal assemblage is a mathematical object that carries the information of all possible correlations that the uncharacterized parties (Bob and Charlie) could generate in a definite causal manner, without even assuming that their operations are restricted by quantum mechanics. See Ref.~\cite{bavaresco19} or App.~\ref{app::SDPcausalmodels} for the precise definition of semi-device-independent causal models.

We now define semi-device-independent causal inequalities, whose violation witnesses the fact that the experimental data cannot be explained by a semi-device-independent causal model. Given a set of experimentally measured correlations $\{p(abc|xyz)\}$ and a description for the operations of Alice $\{\bar{A}_{a|x}\}$, the existence of a causal model that describes this experiment is a membership problem that can be solved via semidefinite programming (SDP)~\cite{boyd04,bavaresco19}. Should the experimental data \textit{not be able} to certify an ICO, this SDP provides us with the exact causal model that describes the data. Alternatively, should the experimental data \textit{be able} to certify ICO, it is guaranteed that there exists a hyperplane that separates the experimental data from the set of correlations that can be described by semi-device-independent causal models. This hyperplane can be obtained by the solution of the dual problem associated to the membership (also called the primal) SDP, and used to construct an inequality of the form
\begin{equation}\label{eq::inequality}
    S \coloneqq \sum_{\substack{abc\\xyz}}\,\alpha^{abc}_{xyz}\,p(abc|xyz) \geq 0,
\end{equation}
where $\{\alpha^{abc}_{xyz}\}$ is a set of real coefficients obtained as the solution of the dual problem. We show that this inequality is satisfied if and only if the data comes from an experiment that: (i) implements a process that is causally separable and (ii) implements the specific instruments $\{\bar{A}_{a|x}\}$, regardless of what operations are performed by Bob and Charlie. Therefore, the violation of this inequality implies that whichever process is being analyzed demonstrates ICO as long as one can guarantee that the hypothesis that the exact instruments $\{\bar{A}_{a|x}\}$ were implemented by Alice holds. The derivation of Ineq.~\eqref{eq::inequality}, as well as the primal and dual problems, can be found in App.~\ref{app::SDPcausalmodels}.

Any set of coefficients $\{\alpha^{abc}_{xyz}\}$ that satisfy the constraints of the dual problem define a valid inequality of the form of Ineq.~\eqref{eq::inequality}. However, when the information of the theoretical prediction of the correlations generated in the experiment is available, our method allows one to derive a specific inequality that is tailored to a particular experiment and able to better unveil the noncausal properties of the process being implemented. In order to derive an inequality tailored to our experiment, we first calculate the theoretically expected set of probability distributions $\{p_\text{theory}(abc|xyz)\}$, using the instruments and quantum switch process provided in App.~\ref{app::choioperators}. Then, by computationally evaluating the dual SDP problem for this set of theoretical probability distributions and the proposed operations of Alice, we obtain the coefficients $\{\alpha^{abc}_{xyz}\}$, available at the repository in Ref.~\cite{github_bavaresco}. From these coefficients, we calculate an expected theoretical value of $S_\text{theory}=-0.0794$ for a perfect quantum switch.

\subsection{Experimental results with a photonic quantum switch}

\begin{figure*}
\begin{center}
    \includegraphics[width=\textwidth]{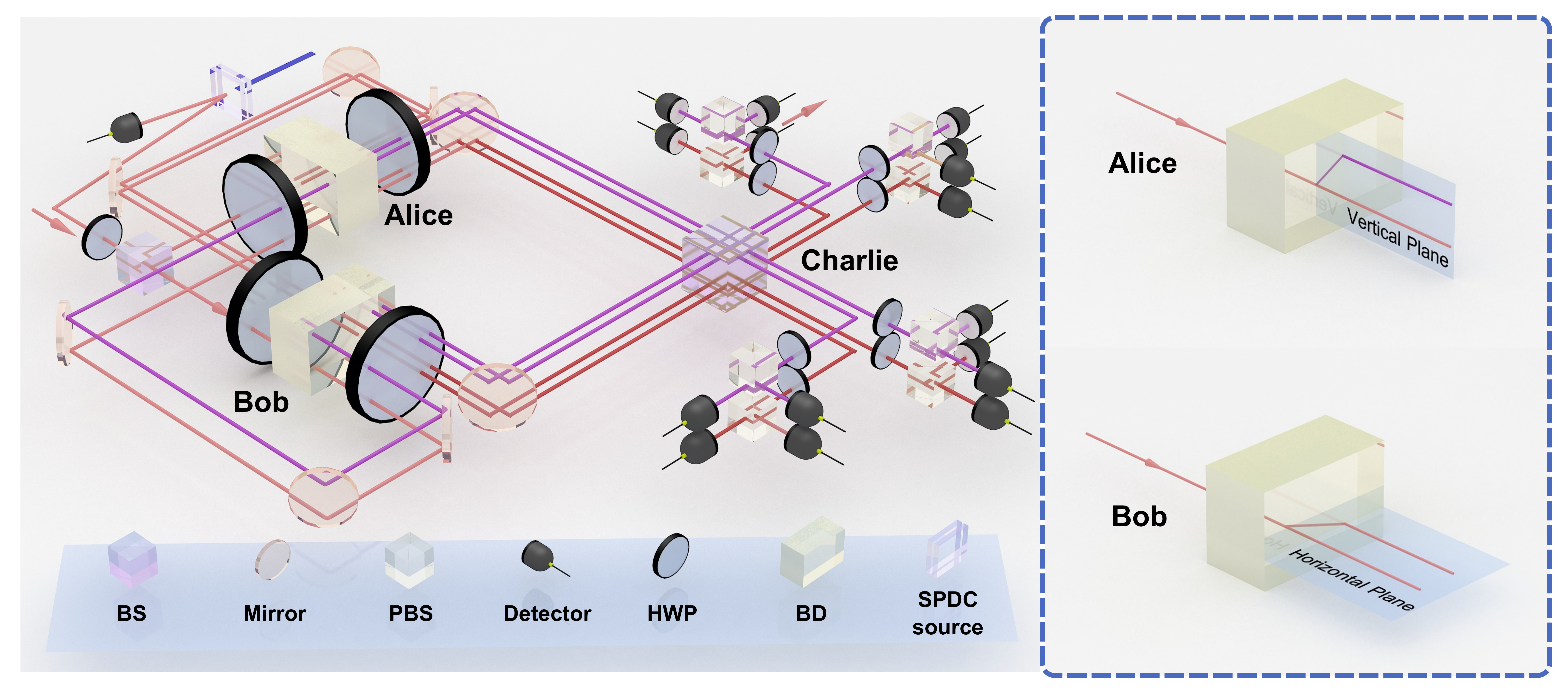}
    \caption{\textbf{Experimental setup.} A 390nm violet laser pumps  $\beta$-barium borate (BBO) crystals cut for beam-like emission to generate a 780nm heralded single photon. Notice that the four interferometer loops form a stereoscopic $2\times2$ optical path array, corresponding to the various outcomes of Alice and Bob's instruments. 
    Bob's outcomes are differentiated by the optical paths in the lower or upper layers (denoted by red and purple beams), while Alice's outcomes are differentiated by the optical paths in the right or left layers.}
    \label{fig:setup}
\end{center}    
\end{figure*}

To experimentally test this semi-device-independent causal inequality, we devised and carried out  {a photonic} quantum switch experiment in which all three involved parties are able to implement multiple-outcome instruments. The experiment starts with the preparation of a heralded single-photon. As shown in Fig.~\ref{fig:setup}, twin photons are generated by spontaneous parametric down-conversion (SPDC). While one is directly detected as a trigger, the heralded signal photon goes through a half-wave plate (HWP) for the initial state preparation and is then fed into a photonic quantum switch. The target qubit is encoded in the polarization of the signal photon, and the control qubit in its path degree of freedom. The path (control) qubit is introduced by the first beam splitter (BS) and the superposition of causal orders is completed when the paths are coherently combined as in a Mach-Zehnder (MZ) interferometer at the second BS, projecting the path qubit into the diagonal basis $\{\ket{\pm}\}$. {Note that in all photonic quantum switches proposed or implemented to date, the operations of Alice and Bob must act identically on two orthogonal optical modes. In most cases, as in our experiment, these modes are two spatial modes that traverse the same optical element, but polarization modes~\cite{goswami18} have been demonstrated, and temporal modes have been proposed~\cite{rambo2016functional}}.

The core of our experimental implementation is the incorporation of multiple-outcome instruments acting on the quantum switch. Only one experiment so far generated local outcomes for two of the three parties of the quantum switch, by coherently adding the outputs of measure-and-reprepare instruments of a single party with two interferometer loops~\cite{rubino17}. Here, our experiment is based on a novel design of a compact setup that allows the incorporation of multiple-outcome instruments for all three parties. Specifically, Alice and Bob perform two different two-outcome measure-and-reprepare instruments on the target system and Charlie performs two different four-outcome projective measurements on target and control systems. Overall, this constitutes $8$ joint input settings and $16$ possible joint outcome sets. 

\begin{figure*}
\begin{center}
    \includegraphics[width=\textwidth]{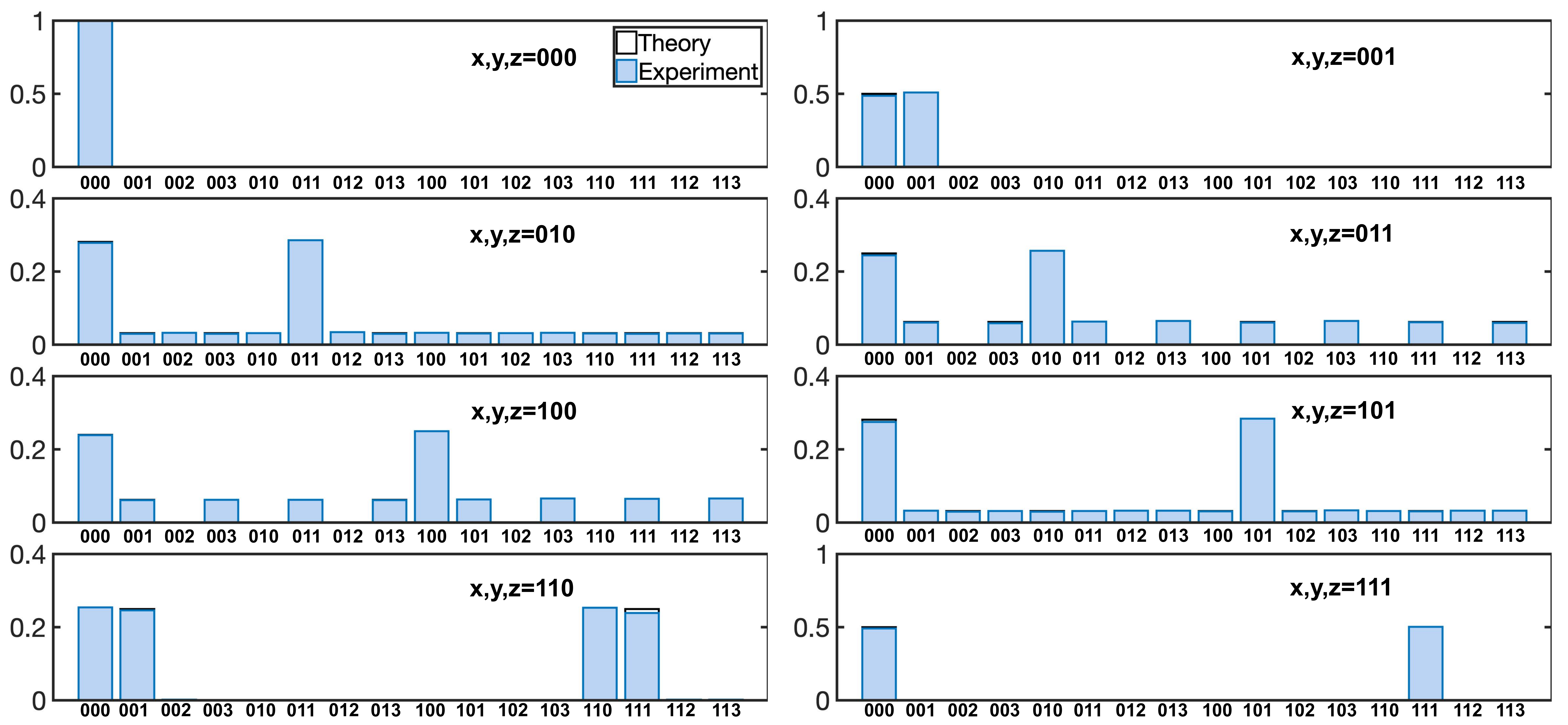}
    \caption{\textbf{Theoretically predicted and experimentally measured frequencies.} Each subplot corresponds to the distribution over one set of joint outcomes, denoted in the horizontal axis by the labels $\{a,b,c\}$, of one input setting, denoted in the title of the subplots by the labels $\{x,y,z\}$. The experimental data is plotted with blue-colored pillars, while the theoretical predictions are plotted with transparent pillars. For each subplot, the probability distribution is normalized.}
    \label{fig:result}
\end{center}    
\end{figure*}

Although measure-and-reprepare instruments can lead to a stronger certification of ICO by allowing one not to make assumptions about the instruments of Bob and Charlie, they also impose the experimental challenge of collecting the outcome statistics generated by Alice and Bob without destroying the coherent superposition of causal orders. We overcome this challenge in the following way: the measure-and-reprepare instruments of Alice and Bob are realized by coupling the polarization mode to additional spatial modes. The measurement step is realized by a HWP followed by a beam displacer (BD), and a subsequent HWP that reprepares the target qubit into the state corresponding to the measurement outcome. The deflecting direction of the BDs of the two parties is set to be orthogonal. Alice's instruments deflect her outcomes horizontally,  while Bob's instruments deflect his outcomes vertically (Fig.~\ref{fig:setup} insets). Consequently, the interferometer loops introduced by the outcome sets $\left(a,b \right)$ of both parties constitute a $2\times2$ interferometer array, with the beams in the lower and upper layers represented in Fig.~\ref{fig:setup} by red and purple beams, respectively. These interferometric loops are introduced to coherently recombine the different spatial modes in order to erase the information about the path through which they propagated. In this way, Alice and Bob operate on the target system locally but do not read out their classical outcomes until the information about the order of their operations is erased, preserving a coherent superposition of causal orders.

The compact interferometer loops are spatially close, in such a way that they undergo essentially the same environmental disturbance, which is inherently nearly phase-synchronous. Hence, they can be simultaneously stabilized with a single phase-locking system. An additional reference beam (not shown in Fig.~\ref{fig:setup}) is fed into the quantum switch and an active locking system is applied to simultaneously lock the phase of all four interferometers (see Materials and Methods). Due to the compact setup and the active locking, an average MZ interference visibility of $>99.1\%$ is achieved for all four interferometer loops over more than 1800 seconds (see Materials and Methods).

To ensure that the device-dependent assumption of Alice's operations holds, we performed quantum process tomography of her local instruments. Fidelities of $>99.8\%$ for all instrument elements of Alice's instruments are achieved, confirming that Alice performs the assumed operations (see App.~\ref{app::tomography}). We recall that the operations of Bob and Charlie are experimentally constructed with the aim of implementing a specific set of operations that is known to allow for the certification of an ICO. However, this information is not taken into account in the analysis of the data. Therefore, our certification does not depend on the implementation of Bob and Charlie's operations being accurate or even in any way close to what is theoretically proposed.

Figure~\ref{fig:result} displays the joint probability distributions over the outcomes of all combinations of instruments performed by each party, both the theoretical prediction and the experimentally collected data. From this experimental data, we achieve a value of $S_\text{exp}=-0.0673 \pm 0.0003 $. The uncertainty of $S_\text{exp}$ represents a single standard deviation via 50 samples of Poisson-distributed photon counts generated by a Monte Carlo simulation. This constitutes a violation by more than $224$ standard deviations.


\section{Discussion}

While a fully device-independent certification is the ultimate goal in the demonstration of  indefinite causal order, currently it is unknown whether the required processes can be physically implemented. Therefore, in the state-of-the-art of quantum process implementation~\cite{wechs21}, the strongest possible method to experimentally certify indefinite causal order is the one that relies on the fewest possible number of device-characterization assumptions. In that sense, we have improved upon previous experimental demonstrations by requiring that only the quantum operations of a single party are characterized. Our proof was based on the conclusive violation of what we introduce as a semi-device-independent causal inequality. Since the quantum switch is known not to produce correlations that can be fully device-independently observed~\cite{araujo15}, our technique constitutes the minimal set of complete device-characterization assumptions necessary for the certification of the non-causal properties of the quantum switch~\cite{bavaresco19}. We ensured these minimal assumptions to hold by performing process tomography of the characterized party. Our setup was based on a novel design of a photonic quantum switch with compact interferometer arrays, that enabled the additional experimental improvement of locally generating multiple outcomes by all parties, instead of the more usual implementation of local (single-outcome) unitary operations. Our active phase-locking system created a very stable and high-performance quantum switch. This demonstration therefore contributes with stronger experimental evidence of the occurrence of indefinite causal order. 

Although our experiment treats all but one party device independently, our demonstration could still be susceptible to loopholes. For example, a quantum switch that acts solely on unitary operations is susceptible to the many-copy loophole~\cite{chiribella13switch}, i.e., this switch could be simulated by a causally ordered circuit that has access to at least one more copy of said unitary operations. To avoid this loophole, in principle, the number of uses of each operation should be certified.
It has been proposed that this could be achieved with a counter device \cite{araujo2014computational,oreshkov2019time}, or by quantifying how much energy has been expended in a single run of the experiment \cite{fellous2022comparing}.
However, it is currently unknown whether a switch that acts on non-unitary operations, such as the ones we implement here, could also be simulated by a circuit with access to more copies of these operations. Hence, it is unclear whether our experiment in particular also suffers from the many-copy loophole.
This current lack of understanding illustrates that the certification of indefinite causal order is a field in its infancy, and more work is required to identify all potential loopholes and devise methods to close them. We hope that our work will further motivate the study of potential loopholes in general quantum switches and that our novel method of implementing operations beyond unitaries will motivate the use of more complex operations in future protocols concerning indefinite causality, stimulating further investigation of this phenomenon based on even less assumptions.


\section{MATERIALS AND METHODS}

\subsection{Phase-locking system setup}

The violation of our semi-device-independent causal inequality relies on integrating measure-and-reprepare operations in both interrogating agents inside the quantum switch. This particular demand can be well addressed in our optical quantum switch with compact interferometer loops. Although the interferometer loops of $2\times2$ interferometer arrays are inherently phase synchronous, they still undergo the identical environmental noise. Actively phase locking system is applied to stabilize the path difference of the two spatial paths introduced by the first beam splitter (BS). 

\begin{figure}
\begin{center}
    \includegraphics[width=\columnwidth]{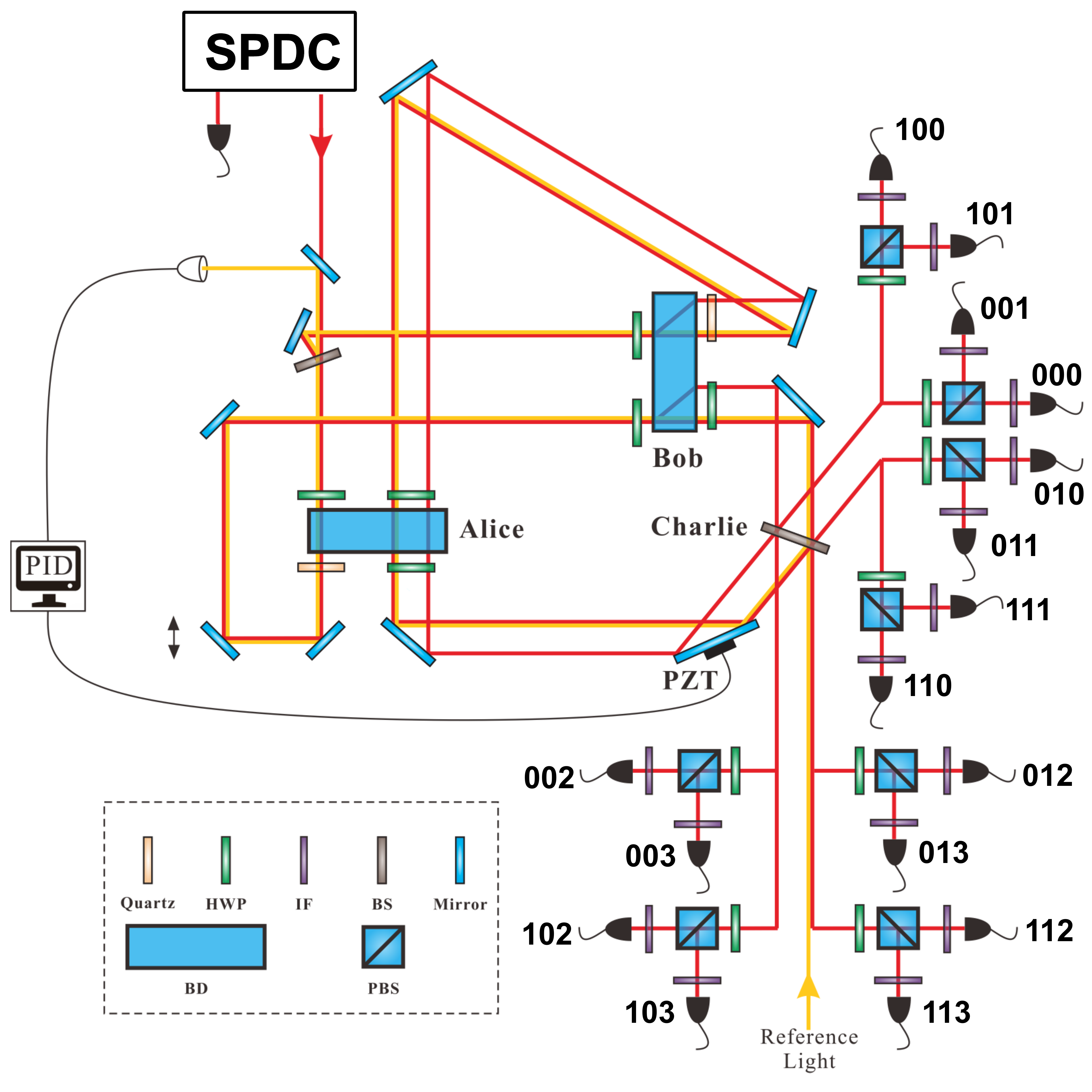}
    \caption{\textbf{Sketch of the experimental setup with locking systems.} PID: proportional–integral–derivative controller; PZT: piezoelectric ceramic; HWP: half-wave plate; IF: interference filter; BS: beam splitter; BD: beam displacer; PBS: polarized beam splitter. Here, the lower-layer and upper-layer interferometers are implicitly shown in a single path. An explicit structure of the interferometer loops is shown in the main text. The single-photon target system is denoted by the red line and the reference light for the locking system is represented in yellow. Two mirrors in the left-bottom part of the figure, mounted in translation stages, work as a trombone-arm delay line to tune the path difference of the two branches of the Mach-Zehnder interferometer (MZI). With the observed interference, the quartz plates are used to independently finely tune the phase of each interferometer loop. Notice that Alice and Bob perform multiple-outcome instruments inside the switch, but do not record the result to avoid destroying the indistinguishability of causal orders until Charlie has performed his measurements. The label in the final single-photon detectors $\{000,001,002,003 \cdots, 110,111,112,113\}$ denotes the joint measurement outcomes $\{a\,b\,c\}_{a,b,c}$ of Alice, Bob and Charlie, respectively. Note that Charlie's instruments include both the measurement on the control qubit (by the BS) and the subsequent measurement on the target qubit (by the polarization analysis system).}
    \label{fig:lockingsetup}
\end{center}    
\end{figure}

The phase locking system consists of a reference light with 780 nm, a photon detector (PD) and a proportional–integral–derivative controller (PID) module. As shown in Fig.~\ref{fig:lockingsetup}, the vertical polarized reference light (the yellow line) is first fed into the quantum switch from its outcome path and counter-propagated through the spatial path. Notice that the reference light is set slightly higher than, but close to, the single photon's optical path, so that it can be conveniently fed into, and separated from, the experimental setup while undergoing the same environmental noise. The reference light is split into two branches by the BS of Charlies' instrument. Each branch travels along its corresponding arm of the Mach-Zehnder interferometer (MZI). The reference light is kept overhead of all the wave plates of Alice and Bob's instrument and only travels through the beam displacers (BDs), such that the polarization is always kept in vertical direction and not deflected by the BD when Alice and Bob run over all settings. The two branches of the MZI are recombined again in first BS. One of the outcomes is reflected by a mirror overhead and continuously monitored by a photon detector (PD). The power recorded by the PD reveals the phase relation of the MZI and is sent into the PID. The feedback voltage generated by PID drives the piezoelectric ceramic (PZT) attached to the mirror to actively stabilize the phase of the MZI. \\ 

\subsection{Performance of interferometer array}

When the locking system is applied, the phases of the four interferometer loops can be efficiently stabilized, while still not sharing precisely the same phase relations because they do not strictly overlap in space. To further finely tune the phases and make them strictly synchronous, additional quartz plates are inserted in each branch of the interferometer. As shown in Fig.~\ref{fig:lockingsetup}, for each spatial branch of the interferometer, one quartz is inserted behind the BD. The optical axes of the quartz are both set along horizontally, such that the quartz will introduce birefringence between horizontal and vertical polarization. By tuning the yaw of the quartz, its effective thickness can be changed so that to introduce a tunable relative phase between outcome spatial modes of BD. The quartz in Alice's side is used to synchronize the phase between the lower (vertical polarization) and upper (horizontal polarization) interferometers (in Fig.~\ref{fig:lockingsetup} they are represented by a single optical line as overlapped in a bird-eye view); The quartz in Bob's side is used to synchronize the phase between the left and right interferometers. 

\begin{figure}
\begin{center}
    \includegraphics[width=\columnwidth]{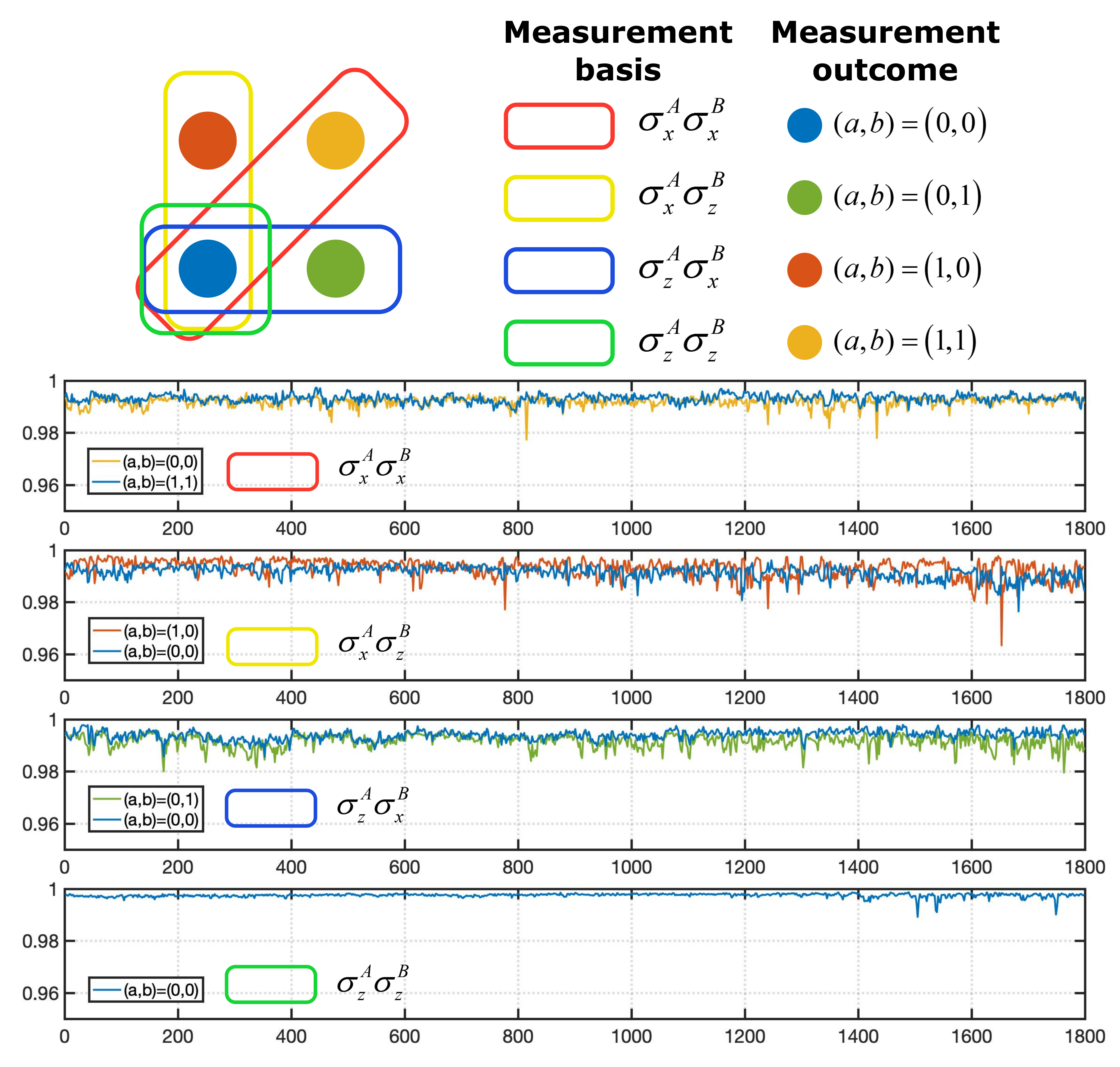}
    \caption{\textbf{Visibilities of the interferometer array.} For each specific setting, only the parts of the spatial mode containing interference are shown. In the upper inset, the colored boxes include the interfering spatial modes. Different colored boxes correspond to different settings, while different colored spatial modes correspond to different measurement outcomes. The interference fringes of these interfering spatial modes are provided in the lower plots, where each setting is plotted individually.}
    \label{fig:visibility}
\end{center}    
\end{figure}

The $2\times2$ interferometer loops will share stabilized and identical phases after the above-mentioned procedures are completed. To characterize the performance of our quantum switch, we measure the interfering visibilities of the $2\times2$ arrays with a single photon in a real experimental scenario. In this real experimental scenario, Alice and Bob may have two possible measurement settings along the direction of $\sigma_{x}$ and $\sigma_{z}$. There are four possible setting combinations in the experiment $\{ \sigma^{A}_{x}\sigma^{B}_{x},\sigma^{A}_{x}\sigma^{B}_{z},\sigma^{A}_{z}\sigma^{B}_{x},\sigma^{A}_{z}\sigma^{B}_{z}\}$, in each setting, there are at most two interferometer loops containing interference, while the others have no interference because in such setting those outcomes may have vanishing probability. In the upper inset of Fig.~\ref{fig:visibility}, we provide the cross-section of the interferometer array and interconnect the spatial modes with corresponding measurement outcomes by different colors. The interferometer loops that contain interference are also 
included by different color boxes for different settings. The lower plots of Fig.~\ref{fig:visibility} depict the measured visibilities of these interferometer loops. Visibilities for each setting are plotted in an individual subplot and each interfering outcome is plotted with a corresponding colored fringe in the upper inset. All fringes show averaged visibility of more than $99.1\%$ in more than 1800s, which suffices to collect the data of the whole experiment. The high visibilities of the interferometer arrays suggest a strong proof that we faithfully implemented a high-performance quantum switch which incorporates distinct outcomes in both Alice's and Bob's side.
\\

\noindent\textbf{Funding.} JB acknowledges the Austrian Science Fund (FWF) through the Zukunftskolleg project ZK03 and the Swiss National Science Foundation (NCCR SwissMAP). This work was supported by the National Key Research and Development Program of China (No. 2017YFA0304100, No. 2016YFA0301300 and No. 2016YFA0301700), the Innovation Program for Quantum Science and Technology (No. 2021ZD0301604), National Natural Science Foundation of China (Nos. 11774335, 11734015, 11874345, 11821404, 11904357); the Austrian Science Fund (FWF): F 7113-N38 (BeyondC), FG 5-N (Research Group); Research Platform for Testing the Quantum and Gravity Interface (TURIS), the European Commission (ErBeSta (No.800942)), Christian Doppler Forschungsgesellschaft;  {\"O}sterreichische Nationalstiftung f{\"u}r Forschung, Technologie und Entwicklung;  Bundesministerium f{\"u}r Digitalisierung und Wirtschaftsstandort.
\\

\noindent\textbf{Acknowledgements.} We are grateful to M.T. Quintino for insightful discussions and comments on the manuscript.
\\

\noindent\textbf{Disclosure.} The authors declare no conflicts of interest.
\\

\noindent\textbf{Data availability.} The code developed for this work is made available by JB at \href{https://github.com/jessicabavaresco/experimental-SDI-causality}{https://github.com/jessicabavaresco/experimental-SDI-causality}. All relevant data is presented in the paper and/or its appendix. Additional data related to this paper may be requested to the authors.


\begin{thebibliography}{52}%
\makeatletter
\providecommand \@ifxundefined [1]{%
 \@ifx{#1\undefined}
}%
\providecommand \@ifnum [1]{%
 \ifnum #1\expandafter \@firstoftwo
 \else \expandafter \@secondoftwo
 \fi
}%
\providecommand \@ifx [1]{%
 \ifx #1\expandafter \@firstoftwo
 \else \expandafter \@secondoftwo
 \fi
}%
\providecommand \natexlab [1]{#1}%
\providecommand \enquote  [1]{``#1''}%
\providecommand \bibnamefont  [1]{#1}%
\providecommand \bibfnamefont [1]{#1}%
\providecommand \citenamefont [1]{#1}%
\providecommand \href@noop [0]{\@secondoftwo}%
\providecommand \href [0]{\begingroup \@sanitize@url \@href}%
\providecommand \@href[1]{\@@startlink{#1}\@@href}%
\providecommand \@@href[1]{\endgroup#1\@@endlink}%
\providecommand \@sanitize@url [0]{\catcode `\\12\catcode `\$12\catcode
  `\&12\catcode `\#12\catcode `\^12\catcode `\_12\catcode `\%12\relax}%
\providecommand \@@startlink[1]{}%
\providecommand \@@endlink[0]{}%
\providecommand \url  [0]{\begingroup\@sanitize@url \@url }%
\providecommand \@url [1]{\endgroup\@href {#1}{\urlprefix }}%
\providecommand \urlprefix  [0]{URL }%
\providecommand \Eprint [0]{\href }%
\providecommand \doibase [0]{https://doi.org/}%
\providecommand \selectlanguage [0]{\@gobble}%
\providecommand \bibinfo  [0]{\@secondoftwo}%
\providecommand \bibfield  [0]{\@secondoftwo}%
\providecommand \translation [1]{[#1]}%
\providecommand \BibitemOpen [0]{}%
\providecommand \bibitemStop [0]{}%
\providecommand \bibitemNoStop [0]{.\EOS\space}%
\providecommand \EOS [0]{\spacefactor3000\relax}%
\providecommand \BibitemShut  [1]{\csname bibitem#1\endcsname}%
\let\auto@bib@innerbib\@empty
\bibitem [{\citenamefont {Bell}(1964)}]{bell64}%
  \BibitemOpen
  \bibfield  {author} {\bibinfo {author} {\bibfnamefont {J.~S.}\ \bibnamefont
  {Bell}},\ }\bibfield  {title} {\bibinfo {title} {{On the Einstein Podolsky
  Rosen paradox}},\ }\href
  {https://doi.org/10.1103/PhysicsPhysiqueFizika.1.195} {\bibfield  {journal}
  {\bibinfo  {journal} {Physics Physique Fizika}\ }\textbf {\bibinfo {volume}
  {1}},\ \bibinfo {pages} {195} (\bibinfo {year} {1964})}\BibitemShut {NoStop}%
\bibitem [{\citenamefont {Chiribella}\ \emph {et~al.}(2013)\citenamefont
  {Chiribella}, \citenamefont {D'Ariano}, \citenamefont {Perinotti},\ and\
  \citenamefont {Valiron}}]{chiribella13switch}%
  \BibitemOpen
  \bibfield  {author} {\bibinfo {author} {\bibfnamefont {G.}~\bibnamefont
  {Chiribella}}, \bibinfo {author} {\bibfnamefont {G.~M.}\ \bibnamefont
  {D'Ariano}}, \bibinfo {author} {\bibfnamefont {P.}~\bibnamefont
  {Perinotti}},\ and\ \bibinfo {author} {\bibfnamefont {B.}~\bibnamefont
  {Valiron}},\ }\bibfield  {title} {\bibinfo {title} {Quantum computations
  without definite causal structure},\ }\href
  {https://doi.org/10.1103/PhysRevA.88.022318} {\bibfield  {journal} {\bibinfo
  {journal} {Phys. Rev. A}\ }\textbf {\bibinfo {volume} {88}},\ \bibinfo
  {pages} {022318} (\bibinfo {year} {2013})},\ \Eprint
  {https://arxiv.org/abs/0912.0195} {arXiv:0912.0195 [quant-ph]} \BibitemShut
  {NoStop}%
\bibitem [{\citenamefont {Oreshkov}\ \emph {et~al.}(2012)\citenamefont
  {Oreshkov}, \citenamefont {Costa},\ and\ \citenamefont
  {Brukner}}]{oreshkov12}%
  \BibitemOpen
  \bibfield  {author} {\bibinfo {author} {\bibfnamefont {O.}~\bibnamefont
  {Oreshkov}}, \bibinfo {author} {\bibfnamefont {F.}~\bibnamefont {Costa}},\
  and\ \bibinfo {author} {\bibfnamefont {{\v{C}}.}~\bibnamefont {Brukner}},\
  }\bibfield  {title} {\bibinfo {title} {Quantum correlations with no causal
  order},\ }\href {https://doi.org/10.1038/ncomms2076} {\bibfield  {journal}
  {\bibinfo  {journal} {Nat. Commun.}\ }\textbf {\bibinfo {volume} {3}},\
  \bibinfo {pages} {1092} (\bibinfo {year} {2012})},\ \Eprint
  {https://arxiv.org/abs/1105.4464} {arXiv:1105.4464 [quant-ph]} \BibitemShut
  {NoStop}%
\bibitem [{\citenamefont {Nielsen}\ and\ \citenamefont
  {Chuang}(2000)}]{nielsenchuang}%
  \BibitemOpen
  \bibfield  {author} {\bibinfo {author} {\bibfnamefont {M.~A.}\ \bibnamefont
  {Nielsen}}\ and\ \bibinfo {author} {\bibfnamefont {I.~L.}\ \bibnamefont
  {Chuang}},\ }\href
  {http://mmrc.amss.cas.cn/tlb/201702/W020170224608149940643.pdf} {\emph
  {\bibinfo {title} {Quantum Computation and Quantum Information}}}\ (\bibinfo
  {publisher} {Cambridge University Press},\ \bibinfo {year}
  {2000})\BibitemShut {NoStop}%
\bibitem [{\citenamefont {Chiribella}\ \emph {et~al.}(2008)\citenamefont
  {Chiribella}, \citenamefont {D'Ariano},\ and\ \citenamefont
  {Perinotti}}]{chiribella08architecture}%
  \BibitemOpen
  \bibfield  {author} {\bibinfo {author} {\bibfnamefont {G.}~\bibnamefont
  {Chiribella}}, \bibinfo {author} {\bibfnamefont {G.~M.}\ \bibnamefont
  {D'Ariano}},\ and\ \bibinfo {author} {\bibfnamefont {P.}~\bibnamefont
  {Perinotti}},\ }\bibfield  {title} {\bibinfo {title} {Quantum circuit
  architecture},\ }\href {https://doi.org/10.1103/PhysRevLett.101.060401}
  {\bibfield  {journal} {\bibinfo  {journal} {Phys. Rev. Lett.}\ }\textbf
  {\bibinfo {volume} {101}},\ \bibinfo {pages} {060401} (\bibinfo {year}
  {2008})},\ \Eprint {https://arxiv.org/abs/0712.1325} {arXiv:0712.1325
  [quant-ph]} \BibitemShut {NoStop}%
\bibitem [{\citenamefont {Kretschmann}\ and\ \citenamefont
  {Werner}(2005)}]{kretschmann05}%
  \BibitemOpen
  \bibfield  {author} {\bibinfo {author} {\bibfnamefont {D.}~\bibnamefont
  {Kretschmann}}\ and\ \bibinfo {author} {\bibfnamefont {R.~F.}\ \bibnamefont
  {Werner}},\ }\bibfield  {title} {\bibinfo {title} {Quantum channels with
  memory},\ }\href {https://doi.org/10.1103/PhysRevA.72.062323} {\bibfield
  {journal} {\bibinfo  {journal} {Phys. Rev. A}\ }\textbf {\bibinfo {volume}
  {72}},\ \bibinfo {pages} {062323} (\bibinfo {year} {2005})},\ \Eprint
  {https://arxiv.org/abs/quant-ph/0502106} {arXiv:quant-ph/0502106}
  \BibitemShut {NoStop}%
\bibitem [{\citenamefont {Gutoski}\ and\ \citenamefont
  {Watrous}(2007)}]{gustoki06}%
  \BibitemOpen
  \bibfield  {author} {\bibinfo {author} {\bibfnamefont {G.}~\bibnamefont
  {Gutoski}}\ and\ \bibinfo {author} {\bibfnamefont {J.}~\bibnamefont
  {Watrous}},\ }\bibfield  {title} {\bibinfo {title} {Toward a general theory
  of quantum games},\ }in\ \href {https://doi.org/10.1145/1250790.1250873}
  {\emph {\bibinfo {booktitle} {Proceedings of the 39th Annual ACM STOC '07}}}\
  (\bibinfo  {publisher} {Association for Computing Machinery},\ \bibinfo
  {address} {New York, NY, USA},\ \bibinfo {year} {2007})\ pp.\ \bibinfo
  {pages} {565--574},\ \Eprint {https://arxiv.org/abs/quant-ph/0611234}
  {arXiv:quant-ph/0611234} \BibitemShut {NoStop}%
\bibitem [{\citenamefont {Chiribella}\ \emph {et~al.}(2009)\citenamefont
  {Chiribella}, \citenamefont {D'Ariano},\ and\ \citenamefont
  {Perinotti}}]{chiribella09}%
  \BibitemOpen
  \bibfield  {author} {\bibinfo {author} {\bibfnamefont {G.}~\bibnamefont
  {Chiribella}}, \bibinfo {author} {\bibfnamefont {G.~M.}\ \bibnamefont
  {D'Ariano}},\ and\ \bibinfo {author} {\bibfnamefont {P.}~\bibnamefont
  {Perinotti}},\ }\bibfield  {title} {\bibinfo {title} {Theoretical framework
  for quantum networks},\ }\href {https://doi.org/10.1103/PhysRevA.80.022339}
  {\bibfield  {journal} {\bibinfo  {journal} {Phys. Rev. A}\ }\textbf {\bibinfo
  {volume} {80}},\ \bibinfo {pages} {022339} (\bibinfo {year} {2009})},\
  \Eprint {https://arxiv.org/abs/0904.4483} {arXiv:0904.4483 [quant-ph]}
  \BibitemShut {NoStop}%
\bibitem [{\citenamefont {Pollock}\ \emph {et~al.}(2018)\citenamefont
  {Pollock}, \citenamefont {Rodr{\'\i}guez-Rosario}, \citenamefont
  {Frauenheim}, \citenamefont {Paternostro},\ and\ \citenamefont
  {Modi}}]{pollock18}%
  \BibitemOpen
  \bibfield  {author} {\bibinfo {author} {\bibfnamefont {F.~A.}\ \bibnamefont
  {Pollock}}, \bibinfo {author} {\bibfnamefont {C.}~\bibnamefont
  {Rodr{\'\i}guez-Rosario}}, \bibinfo {author} {\bibfnamefont {T.}~\bibnamefont
  {Frauenheim}}, \bibinfo {author} {\bibfnamefont {M.}~\bibnamefont
  {Paternostro}},\ and\ \bibinfo {author} {\bibfnamefont {K.}~\bibnamefont
  {Modi}},\ }\bibfield  {title} {\bibinfo {title} {Non-markovian quantum
  processes: Complete framework and efficient characterization},\ }\href
  {https://doi.org/10.1145/1250790.1250873} {\bibfield  {journal} {\bibinfo
  {journal} {Phys. Rev. A}\ }\textbf {\bibinfo {volume} {97}},\ \bibinfo
  {pages} {012127} (\bibinfo {year} {2018})},\ \Eprint
  {https://arxiv.org/abs/1512.00589} {arXiv:1512.00589 [quant-ph]} \BibitemShut
  {NoStop}%
\bibitem [{\citenamefont {Chiribella}(2012)}]{chiribella12}%
  \BibitemOpen
  \bibfield  {author} {\bibinfo {author} {\bibfnamefont {G.}~\bibnamefont
  {Chiribella}},\ }\bibfield  {title} {\bibinfo {title} {Perfect discrimination
  of no-signalling channels via quantum superposition of causal structures},\
  }\href {https://doi.org/10.1103/PhysRevA.86.040301} {\bibfield  {journal}
  {\bibinfo  {journal} {Phys. Rev. A}\ }\textbf {\bibinfo {volume} {86}},\
  \bibinfo {pages} {040301(R)} (\bibinfo {year} {2012})},\ \Eprint
  {https://arxiv.org/abs/1109.5154} {arXiv:1109.5154 [quant-ph]} \BibitemShut
  {NoStop}%
\bibitem [{\citenamefont {Ara{\'{u}}jo}\ \emph
  {et~al.}(2014{\natexlab{a}})\citenamefont {Ara{\'{u}}jo}, \citenamefont
  {Costa},\ and\ \citenamefont {Brukner}}]{araujo14}%
  \BibitemOpen
  \bibfield  {author} {\bibinfo {author} {\bibfnamefont {M.}~\bibnamefont
  {Ara{\'{u}}jo}}, \bibinfo {author} {\bibfnamefont {F.}~\bibnamefont
  {Costa}},\ and\ \bibinfo {author} {\bibfnamefont {{\v{C}}.}~\bibnamefont
  {Brukner}},\ }\bibfield  {title} {\bibinfo {title} {Computational advantage
  from quantum-controlled ordering of gates},\ }\href
  {https://doi.org/10.1103/PhysRevLett.113.250402} {\bibfield  {journal}
  {\bibinfo  {journal} {Phys. Rev. Lett.}\ }\textbf {\bibinfo {volume} {113}},\
  \bibinfo {pages} {250402} (\bibinfo {year} {2014}{\natexlab{a}})},\ \Eprint
  {https://arxiv.org/abs/1401.8127} {arXiv:1401.8127 [quant-ph]} \BibitemShut
  {NoStop}%
\bibitem [{\citenamefont {Feix}\ \emph {et~al.}(2015)\citenamefont {Feix},
  \citenamefont {Ara{\'{u}}jo},\ and\ \citenamefont {Brukner}}]{feix15}%
  \BibitemOpen
  \bibfield  {author} {\bibinfo {author} {\bibfnamefont {A.}~\bibnamefont
  {Feix}}, \bibinfo {author} {\bibfnamefont {M.}~\bibnamefont {Ara{\'{u}}jo}},\
  and\ \bibinfo {author} {\bibfnamefont {{\v{C}}.}~\bibnamefont {Brukner}},\
  }\bibfield  {title} {\bibinfo {title} {Quantum superposition of the order of
  parties as a communication resource},\ }\href
  {https://doi.org/10.1103/PhysRevA.92.052326} {\bibfield  {journal} {\bibinfo
  {journal} {Phys. Rev. A}\ }\textbf {\bibinfo {volume} {92}},\ \bibinfo
  {pages} {052326} (\bibinfo {year} {2015})},\ \Eprint
  {https://arxiv.org/abs/1508.07840} {arXiv:1508.07840 [quant-ph]} \BibitemShut
  {NoStop}%
\bibitem [{\citenamefont {Gu{\'{e}}rin}\ \emph {et~al.}(2016)\citenamefont
  {Gu{\'{e}}rin}, \citenamefont {Feix}, \citenamefont {Ara{\'{u}}jo},\ and\
  \citenamefont {Brukner}}]{guerin16}%
  \BibitemOpen
  \bibfield  {author} {\bibinfo {author} {\bibfnamefont {P.~A.}\ \bibnamefont
  {Gu{\'{e}}rin}}, \bibinfo {author} {\bibfnamefont {A.}~\bibnamefont {Feix}},
  \bibinfo {author} {\bibfnamefont {M.}~\bibnamefont {Ara{\'{u}}jo}},\ and\
  \bibinfo {author} {\bibfnamefont {{\v{C}}.}~\bibnamefont {Brukner}},\
  }\bibfield  {title} {\bibinfo {title} {Exponential communication complexity
  advantage from quantum superposition of the direction of communication},\
  }\href {https://doi.org/10.1103/PhysRevLett.117.100502} {\bibfield  {journal}
  {\bibinfo  {journal} {Phys. Rev. Lett.}\ }\textbf {\bibinfo {volume} {117}},\
  \bibinfo {pages} {100502} (\bibinfo {year} {2016})},\ \Eprint
  {https://arxiv.org/abs/1605.07372} {arXiv:1605.07372 [quant-ph]} \BibitemShut
  {NoStop}%
\bibitem [{\citenamefont {Ebler}\ \emph {et~al.}(2018)\citenamefont {Ebler},
  \citenamefont {Salek},\ and\ \citenamefont {Chiribella}}]{ebler18}%
  \BibitemOpen
  \bibfield  {author} {\bibinfo {author} {\bibfnamefont {D.}~\bibnamefont
  {Ebler}}, \bibinfo {author} {\bibfnamefont {S.}~\bibnamefont {Salek}},\ and\
  \bibinfo {author} {\bibfnamefont {G.}~\bibnamefont {Chiribella}},\ }\bibfield
   {title} {\bibinfo {title} {Enhanced communication with the assistance of
  indefinite causal order},\ }\href
  {https://doi.org/10.1103/PhysRevLett.120.120502} {\bibfield  {journal}
  {\bibinfo  {journal} {Phys. Rev. Lett.}\ }\textbf {\bibinfo {volume} {120}},\
  \bibinfo {pages} {120502} (\bibinfo {year} {2018})},\ \Eprint
  {https://arxiv.org/abs/1711.10165} {arXiv:1711.10165 [quant-ph]} \BibitemShut
  {NoStop}%
\bibitem [{\citenamefont {Quintino}\ \emph {et~al.}(2019)\citenamefont
  {Quintino}, \citenamefont {Dong}, \citenamefont {Shimbo}, \citenamefont
  {Soeda},\ and\ \citenamefont {Murao}}]{quintino18}%
  \BibitemOpen
  \bibfield  {author} {\bibinfo {author} {\bibfnamefont {M.~T.}\ \bibnamefont
  {Quintino}}, \bibinfo {author} {\bibfnamefont {Q.}~\bibnamefont {Dong}},
  \bibinfo {author} {\bibfnamefont {A.}~\bibnamefont {Shimbo}}, \bibinfo
  {author} {\bibfnamefont {A.}~\bibnamefont {Soeda}},\ and\ \bibinfo {author}
  {\bibfnamefont {M.}~\bibnamefont {Murao}},\ }\bibfield  {title} {\bibinfo
  {title} {Reversing unknown quantum transformations: Universal quantum circuit
  for inverting general unitary operations},\ }\href
  {https://doi.org/10.1103/PhysRevLett.123.210502} {\bibfield  {journal}
  {\bibinfo  {journal} {Phys. Rev. Lett.}\ }\textbf {\bibinfo {volume} {123}},\
  \bibinfo {pages} {210502} (\bibinfo {year} {2019})},\ \Eprint
  {https://arxiv.org/abs/1810.06944} {arXiv:1810.06944 [quant-ph]} \BibitemShut
  {NoStop}%
\bibitem [{\citenamefont {Zhao}\ \emph {et~al.}(2020)\citenamefont {Zhao},
  \citenamefont {Yang},\ and\ \citenamefont {Chiribella}}]{zhao20}%
  \BibitemOpen
  \bibfield  {author} {\bibinfo {author} {\bibfnamefont {X.}~\bibnamefont
  {Zhao}}, \bibinfo {author} {\bibfnamefont {Y.}~\bibnamefont {Yang}},\ and\
  \bibinfo {author} {\bibfnamefont {G.}~\bibnamefont {Chiribella}},\ }\bibfield
   {title} {\bibinfo {title} {Quantum metrology with indefinite causal order},\
  }\href {https://doi.org/10.1103/PhysRevLett.124.190503} {\bibfield  {journal}
  {\bibinfo  {journal} {Phys. Rev. Lett.}\ }\textbf {\bibinfo {volume} {124}},\
  \bibinfo {pages} {190503} (\bibinfo {year} {2020})},\ \Eprint
  {https://arxiv.org/abs/1912.02449} {arXiv:1912.02449 [quant-ph]} \BibitemShut
  {NoStop}%
\bibitem [{\citenamefont {Bavaresco}\ \emph {et~al.}(2021)\citenamefont
  {Bavaresco}, \citenamefont {Murao},\ and\ \citenamefont
  {Quintino}}]{bavaresco21a}%
  \BibitemOpen
  \bibfield  {author} {\bibinfo {author} {\bibfnamefont {J.}~\bibnamefont
  {Bavaresco}}, \bibinfo {author} {\bibfnamefont {M.}~\bibnamefont {Murao}},\
  and\ \bibinfo {author} {\bibfnamefont {M.~T.}\ \bibnamefont {Quintino}},\
  }\bibfield  {title} {\bibinfo {title} {Strict hierarchy between parallel,
  sequential, and indefinite-causal-order strategies for channel
  discrimination},\ }\href
  {https://doi.org/https://link.aps.org/doi/10.1103/PhysRevLett.127.200504}
  {\bibfield  {journal} {\bibinfo  {journal} {Phys. Rev. Lett.}\ }\textbf
  {\bibinfo {volume} {127}},\ \bibinfo {pages} {200504} (\bibinfo {year}
  {2021})},\ \Eprint {https://arxiv.org/abs/2011.08300} {arXiv:2011.08300
  [quant-ph]} \BibitemShut {NoStop}%
\bibitem [{\citenamefont {Bavaresco}\ \emph {et~al.}(2022)\citenamefont
  {Bavaresco}, \citenamefont {Murao},\ and\ \citenamefont
  {Quintino}}]{bavaresco21b}%
  \BibitemOpen
  \bibfield  {author} {\bibinfo {author} {\bibfnamefont {J.}~\bibnamefont
  {Bavaresco}}, \bibinfo {author} {\bibfnamefont {M.}~\bibnamefont {Murao}},\
  and\ \bibinfo {author} {\bibfnamefont {M.~T.}\ \bibnamefont {Quintino}},\
  }\bibfield  {title} {\bibinfo {title} {Unitary channel discrimination beyond
  group structures: {Advantages} of sequential and indefinite-causal-order
  strategies},\ }\href {https://doi.org/10.1063/5.0075919} {\bibfield
  {journal} {\bibinfo  {journal} {J. Math. Phys.}\ }\textbf {\bibinfo {volume}
  {63}},\ \bibinfo {pages} {042203} (\bibinfo {year} {2022})},\ \Eprint
  {https://arxiv.org/abs/2105.13369} {arXiv:2105.13369 [quant-ph]} \BibitemShut
  {NoStop}%
\bibitem [{\citenamefont {Renner}\ and\ \citenamefont {\v{C}aslav
  Brukner}(2021)}]{renner21}%
  \BibitemOpen
  \bibfield  {author} {\bibinfo {author} {\bibfnamefont {M.}~\bibnamefont
  {Renner}}\ and\ \bibinfo {author} {\bibnamefont {\v{C}aslav Brukner}},\
  }\bibfield  {title} {\bibinfo {title} {Experimentally feasible computational
  advantage from quantum superposition of gate orders},\ }\href@noop {} {\
  (\bibinfo {year} {2021})},\ \Eprint {https://arxiv.org/abs/2112.14541}
  {arXiv:2112.14541 [quant-ph]} \BibitemShut {NoStop}%
\bibitem [{\citenamefont {Hardy}(2007)}]{hardy2007towards}%
  \BibitemOpen
  \bibfield  {author} {\bibinfo {author} {\bibfnamefont {L.}~\bibnamefont
  {Hardy}},\ }\bibfield  {title} {\bibinfo {title} {Towards quantum gravity: a
  framework for probabilistic theories with non-fixed causal structure},\
  }\href {https://doi.org/10.1088/1751-8113/40/12/S12} {\bibfield  {journal}
  {\bibinfo  {journal} {Journal of Physics A: Mathematical and Theoretical}\
  }\textbf {\bibinfo {volume} {40}},\ \bibinfo {pages} {3081} (\bibinfo {year}
  {2007})},\ \Eprint {https://arxiv.org/abs/gr-qc/0608043}
  {arXiv:gr-qc/0608043} \BibitemShut {NoStop}%
\bibitem [{\citenamefont {Christodoulou}\ and\ \citenamefont
  {Rovelli}(2019)}]{christodoulou2019possibility}%
  \BibitemOpen
  \bibfield  {author} {\bibinfo {author} {\bibfnamefont {M.}~\bibnamefont
  {Christodoulou}}\ and\ \bibinfo {author} {\bibfnamefont {C.}~\bibnamefont
  {Rovelli}},\ }\bibfield  {title} {\bibinfo {title} {On the possibility of
  laboratory evidence for quantum superposition of geometries},\ }\href
  {https://doi.org/10.1016/j.physletb.2019.03.015} {\bibfield  {journal}
  {\bibinfo  {journal} {Phys. Lett. B}\ }\textbf {\bibinfo {volume} {792}},\
  \bibinfo {pages} {64} (\bibinfo {year} {2019})},\ \Eprint
  {https://arxiv.org/abs/1808.05842} {arXiv:1808.05842 [gr-qc]} \BibitemShut
  {NoStop}%
\bibitem [{\citenamefont {Marletto}\ and\ \citenamefont
  {Vedral}(2017)}]{marletto2017gravitationally}%
  \BibitemOpen
  \bibfield  {author} {\bibinfo {author} {\bibfnamefont {C.}~\bibnamefont
  {Marletto}}\ and\ \bibinfo {author} {\bibfnamefont {V.}~\bibnamefont
  {Vedral}},\ }\bibfield  {title} {\bibinfo {title} {Gravitationally induced
  entanglement between two massive particles is sufficient evidence of quantum
  effects in gravity},\ }\href {https://doi.org/10.1103/PhysRevLett.119.240402}
  {\bibfield  {journal} {\bibinfo  {journal} {Phys. Rev. Lett.}\ }\textbf
  {\bibinfo {volume} {119}},\ \bibinfo {pages} {240402} (\bibinfo {year}
  {2017})},\ \Eprint {https://arxiv.org/abs/1707.06036} {arXiv:1707.06036
  [quant-ph]} \BibitemShut {NoStop}%
\bibitem [{\citenamefont {Procopio}\ \emph {et~al.}(2015)\citenamefont
  {Procopio}, \citenamefont {Moqanaki}, \citenamefont {Ara\'{u}jo},
  \citenamefont {Costa}, \citenamefont {Alonso~Calafell}, \citenamefont {Dowd},
  \citenamefont {Hamel}, \citenamefont {Rozema}, \citenamefont {Brukner},\ and\
  \citenamefont {Walther}}]{procopio15}%
  \BibitemOpen
  \bibfield  {author} {\bibinfo {author} {\bibfnamefont {L.~M.}\ \bibnamefont
  {Procopio}}, \bibinfo {author} {\bibfnamefont {A.}~\bibnamefont {Moqanaki}},
  \bibinfo {author} {\bibfnamefont {M.}~\bibnamefont {Ara\'{u}jo}}, \bibinfo
  {author} {\bibfnamefont {F.}~\bibnamefont {Costa}}, \bibinfo {author}
  {\bibfnamefont {I.}~\bibnamefont {Alonso~Calafell}}, \bibinfo {author}
  {\bibfnamefont {E.~G.}\ \bibnamefont {Dowd}}, \bibinfo {author}
  {\bibfnamefont {D.~R.}\ \bibnamefont {Hamel}}, \bibinfo {author}
  {\bibfnamefont {L.~A.}\ \bibnamefont {Rozema}}, \bibinfo {author}
  {\bibfnamefont {{\v{C}}.}~\bibnamefont {Brukner}},\ and\ \bibinfo {author}
  {\bibfnamefont {P.}~\bibnamefont {Walther}},\ }\bibfield  {title} {\bibinfo
  {title} {Experimental superposition of orders of quantum gates},\ }\href
  {https://doi.org/10.1038/ncomms8913} {\bibfield  {journal} {\bibinfo
  {journal} {Nat. Commun.}\ }\textbf {\bibinfo {volume} {6}},\ \bibinfo {pages}
  {7913} (\bibinfo {year} {2015})},\ \Eprint {https://arxiv.org/abs/1412.4006}
  {arXiv:1412.4006 [quant-ph]} \BibitemShut {NoStop}%
\bibitem [{\citenamefont {Rubino}\ \emph {et~al.}(2017)\citenamefont {Rubino},
  \citenamefont {Rozema}, \citenamefont {Feix}, \citenamefont {Ara\'{u}jo},
  \citenamefont {Zeuner}, \citenamefont {Procopio}, \citenamefont {Brukner},\
  and\ \citenamefont {Walther}}]{rubino17}%
  \BibitemOpen
  \bibfield  {author} {\bibinfo {author} {\bibfnamefont {G.}~\bibnamefont
  {Rubino}}, \bibinfo {author} {\bibfnamefont {L.~A.}\ \bibnamefont {Rozema}},
  \bibinfo {author} {\bibfnamefont {A.}~\bibnamefont {Feix}}, \bibinfo {author}
  {\bibfnamefont {M.}~\bibnamefont {Ara\'{u}jo}}, \bibinfo {author}
  {\bibfnamefont {J.~M.}\ \bibnamefont {Zeuner}}, \bibinfo {author}
  {\bibfnamefont {L.~M.}\ \bibnamefont {Procopio}}, \bibinfo {author}
  {\bibfnamefont {{\v{C}}.}~\bibnamefont {Brukner}},\ and\ \bibinfo {author}
  {\bibfnamefont {P.}~\bibnamefont {Walther}},\ }\bibfield  {title} {\bibinfo
  {title} {Experimental verification of an indefinite causal order},\ }\href
  {https://doi.org/10.1126/sciadv.1602589} {\bibfield  {journal} {\bibinfo
  {journal} {Science Advances}\ }\textbf {\bibinfo {volume} {3}},\ \bibinfo
  {pages} {3} (\bibinfo {year} {2017})},\ \Eprint
  {https://arxiv.org/abs/1608.01683} {arXiv:1608.01683 [quant-ph]} \BibitemShut
  {NoStop}%
\bibitem [{\citenamefont {Goswami}\ \emph {et~al.}(2018)\citenamefont
  {Goswami}, \citenamefont {Giarmatzi}, \citenamefont {Kewming}, \citenamefont
  {Costa}, \citenamefont {Branciard}, \citenamefont {Romero},\ and\
  \citenamefont {White}}]{goswami18}%
  \BibitemOpen
  \bibfield  {author} {\bibinfo {author} {\bibfnamefont {K.}~\bibnamefont
  {Goswami}}, \bibinfo {author} {\bibfnamefont {C.}~\bibnamefont {Giarmatzi}},
  \bibinfo {author} {\bibfnamefont {M.}~\bibnamefont {Kewming}}, \bibinfo
  {author} {\bibfnamefont {F.}~\bibnamefont {Costa}}, \bibinfo {author}
  {\bibfnamefont {C.}~\bibnamefont {Branciard}}, \bibinfo {author}
  {\bibfnamefont {J.}~\bibnamefont {Romero}},\ and\ \bibinfo {author}
  {\bibfnamefont {A.~G.}\ \bibnamefont {White}},\ }\bibfield  {title} {\bibinfo
  {title} {Indefinite causal order in a quantum switch},\ }\href
  {https://doi.org/10.1103/PhysRevLett.121.090503} {\bibfield  {journal}
  {\bibinfo  {journal} {Phys. Rev. Lett.}\ }\textbf {\bibinfo {volume} {121}},\
  \bibinfo {pages} {090503} (\bibinfo {year} {2018})},\ \Eprint
  {https://arxiv.org/abs/1803.04302} {arXiv:1803.04302 [quant-ph]} \BibitemShut
  {NoStop}%
\bibitem [{\citenamefont {Rubino}\ \emph {et~al.}(2022)\citenamefont {Rubino},
  \citenamefont {Rozema}, \citenamefont {Massa}, \citenamefont {Ara{\'{u}}jo},
  \citenamefont {Zych}, \citenamefont {Brukner},\ and\ \citenamefont
  {Walther}}]{rubino22}%
  \BibitemOpen
  \bibfield  {author} {\bibinfo {author} {\bibfnamefont {G.}~\bibnamefont
  {Rubino}}, \bibinfo {author} {\bibfnamefont {L.~A.}\ \bibnamefont {Rozema}},
  \bibinfo {author} {\bibfnamefont {F.}~\bibnamefont {Massa}}, \bibinfo
  {author} {\bibfnamefont {M.}~\bibnamefont {Ara{\'{u}}jo}}, \bibinfo {author}
  {\bibfnamefont {M.}~\bibnamefont {Zych}}, \bibinfo {author} {\bibfnamefont
  {{\v{C}}.}~\bibnamefont {Brukner}},\ and\ \bibinfo {author} {\bibfnamefont
  {P.}~\bibnamefont {Walther}},\ }\bibfield  {title} {\bibinfo {title}
  {Experimental entanglement of temporal order},\ }\href
  {https://doi.org/10.22331/q-2022-01-11-621} {\bibfield  {journal} {\bibinfo
  {journal} {{Quantum}}\ }\textbf {\bibinfo {volume} {6}},\ \bibinfo {pages}
  {621} (\bibinfo {year} {2022})},\ \Eprint {https://arxiv.org/abs/1712.06884}
  {arXiv:1712.06884 [quant-ph]} \BibitemShut {NoStop}%
\bibitem [{\citenamefont {Goswami}\ \emph {et~al.}(2020)\citenamefont
  {Goswami}, \citenamefont {Cao}, \citenamefont {Paz-Silva}, \citenamefont
  {Romero},\ and\ \citenamefont {White}}]{goswami20}%
  \BibitemOpen
  \bibfield  {author} {\bibinfo {author} {\bibfnamefont {K.}~\bibnamefont
  {Goswami}}, \bibinfo {author} {\bibfnamefont {Y.}~\bibnamefont {Cao}},
  \bibinfo {author} {\bibfnamefont {G.~A.}\ \bibnamefont {Paz-Silva}}, \bibinfo
  {author} {\bibfnamefont {J.}~\bibnamefont {Romero}},\ and\ \bibinfo {author}
  {\bibfnamefont {A.~G.}\ \bibnamefont {White}},\ }\bibfield  {title} {\bibinfo
  {title} {Increasing communication capacity via superposition of order},\
  }\href {https://doi.org/10.1103/PhysRevResearch.2.033292} {\bibfield
  {journal} {\bibinfo  {journal} {Phys. Rev. Research}\ }\textbf {\bibinfo
  {volume} {2}},\ \bibinfo {pages} {033292} (\bibinfo {year} {2020})},\ \Eprint
  {https://arxiv.org/abs/1807.07383} {arXiv:1807.07383 [quant-ph]} \BibitemShut
  {NoStop}%
\bibitem [{\citenamefont {Wei}\ \emph {et~al.}(2019)\citenamefont {Wei},
  \citenamefont {Tischler}, \citenamefont {Zhao}, \citenamefont {Li},
  \citenamefont {Arrazola}, \citenamefont {Liu}, \citenamefont {Zhang},
  \citenamefont {Li}, \citenamefont {You}, \citenamefont {Wang}, \citenamefont
  {Chen}, \citenamefont {Sanders}, \citenamefont {Zhang}, \citenamefont
  {Pryde}, \citenamefont {Xu},\ and\ \citenamefont {Pan}}]{wei19}%
  \BibitemOpen
  \bibfield  {author} {\bibinfo {author} {\bibfnamefont {K.}~\bibnamefont
  {Wei}}, \bibinfo {author} {\bibfnamefont {N.}~\bibnamefont {Tischler}},
  \bibinfo {author} {\bibfnamefont {S.-R.}\ \bibnamefont {Zhao}}, \bibinfo
  {author} {\bibfnamefont {Y.-H.}\ \bibnamefont {Li}}, \bibinfo {author}
  {\bibfnamefont {J.~M.}\ \bibnamefont {Arrazola}}, \bibinfo {author}
  {\bibfnamefont {Y.}~\bibnamefont {Liu}}, \bibinfo {author} {\bibfnamefont
  {W.}~\bibnamefont {Zhang}}, \bibinfo {author} {\bibfnamefont
  {H.}~\bibnamefont {Li}}, \bibinfo {author} {\bibfnamefont {L.}~\bibnamefont
  {You}}, \bibinfo {author} {\bibfnamefont {Z.}~\bibnamefont {Wang}}, \bibinfo
  {author} {\bibfnamefont {Y.-A.}\ \bibnamefont {Chen}}, \bibinfo {author}
  {\bibfnamefont {B.~C.}\ \bibnamefont {Sanders}}, \bibinfo {author}
  {\bibfnamefont {Q.}~\bibnamefont {Zhang}}, \bibinfo {author} {\bibfnamefont
  {G.~J.}\ \bibnamefont {Pryde}}, \bibinfo {author} {\bibfnamefont
  {F.}~\bibnamefont {Xu}},\ and\ \bibinfo {author} {\bibfnamefont {J.-W.}\
  \bibnamefont {Pan}},\ }\bibfield  {title} {\bibinfo {title} {Experimental
  quantum switching for exponentially superior quantum communication
  complexity},\ }\href {https://doi.org/10.1103/PhysRevLett.122.120504}
  {\bibfield  {journal} {\bibinfo  {journal} {Phys. Rev. Lett.}\ }\textbf
  {\bibinfo {volume} {122}},\ \bibinfo {pages} {120504} (\bibinfo {year}
  {2019})},\ \Eprint {https://arxiv.org/abs/1810.10238} {arXiv:1810.10238
  [quant-ph]} \BibitemShut {NoStop}%
\bibitem [{\citenamefont {Guo}\ \emph {et~al.}(2020)\citenamefont {Guo},
  \citenamefont {Hu}, \citenamefont {Hou}, \citenamefont {Cao}, \citenamefont
  {Cui}, \citenamefont {Liu}, \citenamefont {Huang}, \citenamefont {Li},
  \citenamefont {Guo},\ and\ \citenamefont {Chiribella}}]{guo20}%
  \BibitemOpen
  \bibfield  {author} {\bibinfo {author} {\bibfnamefont {Y.}~\bibnamefont
  {Guo}}, \bibinfo {author} {\bibfnamefont {X.-M.}\ \bibnamefont {Hu}},
  \bibinfo {author} {\bibfnamefont {Z.-B.}\ \bibnamefont {Hou}}, \bibinfo
  {author} {\bibfnamefont {H.}~\bibnamefont {Cao}}, \bibinfo {author}
  {\bibfnamefont {J.-M.}\ \bibnamefont {Cui}}, \bibinfo {author} {\bibfnamefont
  {B.-H.}\ \bibnamefont {Liu}}, \bibinfo {author} {\bibfnamefont {Y.-F.}\
  \bibnamefont {Huang}}, \bibinfo {author} {\bibfnamefont {C.-F.}\ \bibnamefont
  {Li}}, \bibinfo {author} {\bibfnamefont {G.-C.}\ \bibnamefont {Guo}},\ and\
  \bibinfo {author} {\bibfnamefont {G.}~\bibnamefont {Chiribella}},\ }\bibfield
   {title} {\bibinfo {title} {Experimental transmission of quantum information
  using a superposition of causal orders},\ }\href
  {https://doi.org/10.1103/PhysRevLett.124.030502} {\bibfield  {journal}
  {\bibinfo  {journal} {Phys. Rev. Lett.}\ }\textbf {\bibinfo {volume} {124}},\
  \bibinfo {pages} {030502} (\bibinfo {year} {2020})},\ \Eprint
  {https://arxiv.org/abs/1811.07526} {arXiv:1811.07526 [quant-ph]} \BibitemShut
  {NoStop}%
\bibitem [{\citenamefont {Rubino}\ \emph {et~al.}(2021)\citenamefont {Rubino},
  \citenamefont {Rozema}, \citenamefont {Ebler}, \citenamefont
  {Kristj\'ansson}, \citenamefont {Salek}, \citenamefont {Allard~Gu\'erin},
  \citenamefont {Abbott}, \citenamefont {Branciard}, \citenamefont {Brukner},
  \citenamefont {Chiribella},\ and\ \citenamefont {Walther}}]{rubino21}%
  \BibitemOpen
  \bibfield  {author} {\bibinfo {author} {\bibfnamefont {G.}~\bibnamefont
  {Rubino}}, \bibinfo {author} {\bibfnamefont {L.~A.}\ \bibnamefont {Rozema}},
  \bibinfo {author} {\bibfnamefont {D.}~\bibnamefont {Ebler}}, \bibinfo
  {author} {\bibfnamefont {H.}~\bibnamefont {Kristj\'ansson}}, \bibinfo
  {author} {\bibfnamefont {S.}~\bibnamefont {Salek}}, \bibinfo {author}
  {\bibfnamefont {P.}~\bibnamefont {Allard~Gu\'erin}}, \bibinfo {author}
  {\bibfnamefont {A.~A.}\ \bibnamefont {Abbott}}, \bibinfo {author}
  {\bibfnamefont {C.}~\bibnamefont {Branciard}}, \bibinfo {author}
  {\bibfnamefont {i.~c.~v.}\ \bibnamefont {Brukner}}, \bibinfo {author}
  {\bibfnamefont {G.}~\bibnamefont {Chiribella}},\ and\ \bibinfo {author}
  {\bibfnamefont {P.}~\bibnamefont {Walther}},\ }\bibfield  {title} {\bibinfo
  {title} {Experimental quantum communication enhancement by superposing
  trajectories},\ }\href {https://doi.org/10.1103/PhysRevResearch.3.013093}
  {\bibfield  {journal} {\bibinfo  {journal} {Phys. Rev. Research}\ }\textbf
  {\bibinfo {volume} {3}},\ \bibinfo {pages} {013093} (\bibinfo {year}
  {2021})}\BibitemShut {NoStop}%
\bibitem [{\citenamefont {Taddei}\ \emph {et~al.}(2021)\citenamefont {Taddei},
  \citenamefont {Cari\~ne}, \citenamefont {Mart\'{\i}nez}, \citenamefont
  {Garc\'{\i}a}, \citenamefont {Guerrero}, \citenamefont {Abbott},
  \citenamefont {Ara\'ujo}, \citenamefont {Branciard}, \citenamefont {G\'omez},
  \citenamefont {Walborn}, \citenamefont {Aolita},\ and\ \citenamefont
  {Lima}}]{taddei21}%
  \BibitemOpen
  \bibfield  {author} {\bibinfo {author} {\bibfnamefont {M.~M.}\ \bibnamefont
  {Taddei}}, \bibinfo {author} {\bibfnamefont {J.}~\bibnamefont {Cari\~ne}},
  \bibinfo {author} {\bibfnamefont {D.}~\bibnamefont {Mart\'{\i}nez}}, \bibinfo
  {author} {\bibfnamefont {T.}~\bibnamefont {Garc\'{\i}a}}, \bibinfo {author}
  {\bibfnamefont {N.}~\bibnamefont {Guerrero}}, \bibinfo {author}
  {\bibfnamefont {A.~A.}\ \bibnamefont {Abbott}}, \bibinfo {author}
  {\bibfnamefont {M.}~\bibnamefont {Ara\'ujo}}, \bibinfo {author}
  {\bibfnamefont {C.}~\bibnamefont {Branciard}}, \bibinfo {author}
  {\bibfnamefont {E.~S.}\ \bibnamefont {G\'omez}}, \bibinfo {author}
  {\bibfnamefont {S.~P.}\ \bibnamefont {Walborn}}, \bibinfo {author}
  {\bibfnamefont {L.}~\bibnamefont {Aolita}},\ and\ \bibinfo {author}
  {\bibfnamefont {G.}~\bibnamefont {Lima}},\ }\bibfield  {title} {\bibinfo
  {title} {Computational advantage from the quantum superposition of multiple
  temporal orders of photonic gates},\ }\href
  {https://doi.org/10.1103/PRXQuantum.2.010320} {\bibfield  {journal} {\bibinfo
   {journal} {PRX Quantum}\ }\textbf {\bibinfo {volume} {2}},\ \bibinfo {pages}
  {010320} (\bibinfo {year} {2021})},\ \Eprint
  {https://arxiv.org/abs/2002.07817} {arXiv:2002.07817 [quant-ph]} \BibitemShut
  {NoStop}%
\bibitem [{\citenamefont {Paunkovi{\'{c}}}\ and\ \citenamefont
  {Vojinovi{\'{c}}}(2020)}]{paunkovic2020causal}%
  \BibitemOpen
  \bibfield  {author} {\bibinfo {author} {\bibfnamefont {N.}~\bibnamefont
  {Paunkovi{\'{c}}}}\ and\ \bibinfo {author} {\bibfnamefont {M.}~\bibnamefont
  {Vojinovi{\'{c}}}},\ }\bibfield  {title} {\bibinfo {title} {Causal orders,
  quantum circuits and spacetime: distinguishing between definite and
  superposed causal orders},\ }\href
  {https://doi.org/10.22331/q-2020-05-28-275} {\bibfield  {journal} {\bibinfo
  {journal} {{Quantum}}\ }\textbf {\bibinfo {volume} {4}},\ \bibinfo {pages}
  {275} (\bibinfo {year} {2020})},\ \Eprint {https://arxiv.org/abs/1905.09682}
  {arXiv:1905.09682 [quant-ph]} \BibitemShut {NoStop}%
\bibitem [{\citenamefont {Vilasini}\ and\ \citenamefont
  {Renner}(2022)}]{vilasini2022embedding}%
  \BibitemOpen
  \bibfield  {author} {\bibinfo {author} {\bibfnamefont {V.}~\bibnamefont
  {Vilasini}}\ and\ \bibinfo {author} {\bibfnamefont {R.}~\bibnamefont
  {Renner}},\ }\bibfield  {title} {\bibinfo {title} {Embedding cyclic causal
  structures in acyclic spacetimes: no-go results for process matrices},\
  }\href@noop {} {\bibfield  {journal} {\bibinfo  {journal} {arXiv preprint}\ }
  (\bibinfo {year} {2022})},\ \Eprint {https://arxiv.org/abs/2203.11245}
  {arXiv:2203.11245 [quant-ph]} \BibitemShut {NoStop}%
\bibitem [{\citenamefont {Ormrod}\ \emph {et~al.}(2022)\citenamefont {Ormrod},
  \citenamefont {Vanrietvelde},\ and\ \citenamefont
  {Barrett}}]{ormrod2022causal}%
  \BibitemOpen
  \bibfield  {author} {\bibinfo {author} {\bibfnamefont {N.}~\bibnamefont
  {Ormrod}}, \bibinfo {author} {\bibfnamefont {A.}~\bibnamefont
  {Vanrietvelde}},\ and\ \bibinfo {author} {\bibfnamefont {J.}~\bibnamefont
  {Barrett}},\ }\bibfield  {title} {\bibinfo {title} {Causal structure in the
  presence of sectorial constraints, with application to the quantum switch},\
  }\href@noop {} {\bibfield  {journal} {\bibinfo  {journal} {arXiv preprint}\ }
  (\bibinfo {year} {2022})},\ \Eprint {https://arxiv.org/abs/2204.10273}
  {arXiv:2204.10273 [quant-ph]} \BibitemShut {NoStop}%
\bibitem [{\citenamefont {Oreshkov}(2019)}]{oreshkov2019time}%
  \BibitemOpen
  \bibfield  {author} {\bibinfo {author} {\bibfnamefont {O.}~\bibnamefont
  {Oreshkov}},\ }\bibfield  {title} {\bibinfo {title} {Time-delocalized quantum
  subsystems and operations: on the existence of processes with indefinite
  causal structure in quantum mechanics},\ }\href
  {https://doi.org/10.22331/q-2019-12-02-206} {\bibfield  {journal} {\bibinfo
  {journal} {{Quantum}}\ }\textbf {\bibinfo {volume} {3}},\ \bibinfo {pages}
  {206} (\bibinfo {year} {2019})},\ \Eprint {https://arxiv.org/abs/1801.07594}
  {arXiv:1801.07594 [quant-ph]} \BibitemShut {NoStop}%
\bibitem [{\citenamefont {Fellous-Asiani}\ \emph {et~al.}(2022)\citenamefont
  {Fellous-Asiani}, \citenamefont {Mothe}, \citenamefont {Bresque},
  \citenamefont {Dourdent}, \citenamefont {Camati}, \citenamefont {Abbott},
  \citenamefont {Auff{\`e}ves},\ and\ \citenamefont
  {Branciard}}]{fellous2022comparing}%
  \BibitemOpen
  \bibfield  {author} {\bibinfo {author} {\bibfnamefont {M.}~\bibnamefont
  {Fellous-Asiani}}, \bibinfo {author} {\bibfnamefont {R.}~\bibnamefont
  {Mothe}}, \bibinfo {author} {\bibfnamefont {L.}~\bibnamefont {Bresque}},
  \bibinfo {author} {\bibfnamefont {H.}~\bibnamefont {Dourdent}}, \bibinfo
  {author} {\bibfnamefont {P.~A.}\ \bibnamefont {Camati}}, \bibinfo {author}
  {\bibfnamefont {A.~A.}\ \bibnamefont {Abbott}}, \bibinfo {author}
  {\bibfnamefont {A.}~\bibnamefont {Auff{\`e}ves}},\ and\ \bibinfo {author}
  {\bibfnamefont {C.}~\bibnamefont {Branciard}},\ }\bibfield  {title} {\bibinfo
  {title} {Comparing the quantum switch and its simulations with
  energetically-constrained operations},\ }\href@noop {} {\bibfield  {journal}
  {\bibinfo  {journal} {arXiv preprint arXiv:2208.01952}\ } (\bibinfo {year}
  {2022})}\BibitemShut {NoStop}%
\bibitem [{\citenamefont {Branciard}\ \emph {et~al.}(2016)\citenamefont
  {Branciard}, \citenamefont {Ara{\'{u}}jo}, \citenamefont {Feix},
  \citenamefont {Costa},\ and\ \citenamefont {\v{C}. Brukner}}]{branciard16}%
  \BibitemOpen
  \bibfield  {author} {\bibinfo {author} {\bibfnamefont {C.}~\bibnamefont
  {Branciard}}, \bibinfo {author} {\bibfnamefont {M.}~\bibnamefont
  {Ara{\'{u}}jo}}, \bibinfo {author} {\bibfnamefont {A.}~\bibnamefont {Feix}},
  \bibinfo {author} {\bibfnamefont {F.}~\bibnamefont {Costa}},\ and\ \bibinfo
  {author} {\bibnamefont {\v{C}. Brukner}},\ }\bibfield  {title} {\bibinfo
  {title} {The simplest causal inequalities and their violation},\ }\href
  {https://doi.org/10.1088/1367-2630/18/1/013008} {\bibfield  {journal}
  {\bibinfo  {journal} {New J. Phys.}\ }\textbf {\bibinfo {volume} {18}},\
  \bibinfo {pages} {013008} (\bibinfo {year} {2016})},\ \Eprint
  {https://arxiv.org/abs/1508.01704} {arXiv:1508.01704 [quant-ph]} \BibitemShut
  {NoStop}%
\bibitem [{\citenamefont {Brunner}\ \emph {et~al.}(2014)\citenamefont
  {Brunner}, \citenamefont {Cavalcanti}, \citenamefont {Pironio}, \citenamefont
  {Scarani},\ and\ \citenamefont {Wehner}}]{brunner14}%
  \BibitemOpen
  \bibfield  {author} {\bibinfo {author} {\bibfnamefont {N.}~\bibnamefont
  {Brunner}}, \bibinfo {author} {\bibfnamefont {D.}~\bibnamefont {Cavalcanti}},
  \bibinfo {author} {\bibfnamefont {S.}~\bibnamefont {Pironio}}, \bibinfo
  {author} {\bibfnamefont {V.}~\bibnamefont {Scarani}},\ and\ \bibinfo {author}
  {\bibfnamefont {S.}~\bibnamefont {Wehner}},\ }\bibfield  {title} {\bibinfo
  {title} {Bell nonlocality},\ }\href
  {https://doi.org/10.1103/RevModPhys.86.419} {\bibfield  {journal} {\bibinfo
  {journal} {Rev. Mod. Phys.}\ }\textbf {\bibinfo {volume} {86}},\ \bibinfo
  {pages} {419} (\bibinfo {year} {2014})},\ \Eprint
  {https://arxiv.org/abs/1303.2849} {arXiv:1303.2849 [quant-ph]} \BibitemShut
  {NoStop}%
\bibitem [{\citenamefont {Ara{\'{u}}jo}\ \emph {et~al.}(2015)\citenamefont
  {Ara{\'{u}}jo}, \citenamefont {Branciard}, \citenamefont {Costa},
  \citenamefont {Feix}, \citenamefont {Giarmatzi},\ and\ \citenamefont
  {Brukner}}]{araujo15}%
  \BibitemOpen
  \bibfield  {author} {\bibinfo {author} {\bibfnamefont {M.}~\bibnamefont
  {Ara{\'{u}}jo}}, \bibinfo {author} {\bibfnamefont {C.}~\bibnamefont
  {Branciard}}, \bibinfo {author} {\bibfnamefont {F.}~\bibnamefont {Costa}},
  \bibinfo {author} {\bibfnamefont {A.}~\bibnamefont {Feix}}, \bibinfo {author}
  {\bibfnamefont {C.}~\bibnamefont {Giarmatzi}},\ and\ \bibinfo {author}
  {\bibfnamefont {{\v{C}}.}~\bibnamefont {Brukner}},\ }\bibfield  {title}
  {\bibinfo {title} {Witnessing causal nonseparability},\ }\href
  {https://doi.org/10.1088/1367-2630/17/10/102001} {\bibfield  {journal}
  {\bibinfo  {journal} {New J. Phys.}\ }\textbf {\bibinfo {volume} {17}},\
  \bibinfo {pages} {102001} (\bibinfo {year} {2015})},\ \Eprint
  {https://arxiv.org/abs/1506.03776} {arXiv:1506.03776 [quant-ph]} \BibitemShut
  {NoStop}%
\bibitem [{\citenamefont {Terhal}(2000)}]{terhal00}%
  \BibitemOpen
  \bibfield  {author} {\bibinfo {author} {\bibfnamefont {B.~M.}\ \bibnamefont
  {Terhal}},\ }\bibfield  {title} {\bibinfo {title} {Bell inequalities and the
  separability criterion},\ }\href
  {https://doi.org/https://doi.org/10.1016/S0375-9601(00)00401-1} {\bibfield
  {journal} {\bibinfo  {journal} {Phys. Lett. A}\ }\textbf {\bibinfo {volume}
  {271}},\ \bibinfo {pages} {319} (\bibinfo {year} {2000})},\ \Eprint
  {https://arxiv.org/abs/quant-ph/9911057} {arXiv:quant-ph/9911057}
  \BibitemShut {NoStop}%
\bibitem [{\citenamefont {Bavaresco}\ \emph {et~al.}(2019)\citenamefont
  {Bavaresco}, \citenamefont {Ara{\'{u}}jo}, \citenamefont {Brukner},\ and\
  \citenamefont {Quintino}}]{bavaresco19}%
  \BibitemOpen
  \bibfield  {author} {\bibinfo {author} {\bibfnamefont {J.}~\bibnamefont
  {Bavaresco}}, \bibinfo {author} {\bibfnamefont {M.}~\bibnamefont
  {Ara{\'{u}}jo}}, \bibinfo {author} {\bibfnamefont {{\v{C}}.}~\bibnamefont
  {Brukner}},\ and\ \bibinfo {author} {\bibfnamefont {M.~T.}\ \bibnamefont
  {Quintino}},\ }\bibfield  {title} {\bibinfo {title} {Semi-device-independent
  certification of indefinite causal order},\ }\href
  {https://doi.org/10.22331/q-2019-08-19-176} {\bibfield  {journal} {\bibinfo
  {journal} {{Quantum}}\ }\textbf {\bibinfo {volume} {3}},\ \bibinfo {pages}
  {176} (\bibinfo {year} {2019})},\ \Eprint {https://arxiv.org/abs/1903.10526}
  {arXiv:1903.10526 [quant-ph]} \BibitemShut {NoStop}%
\bibitem [{\citenamefont {Dourdent}\ \emph {et~al.}(2022)\citenamefont
  {Dourdent}, \citenamefont {Abbott}, \citenamefont {Brunner}, \citenamefont
  {\ifmmode \check{S}\else \v{S}\fi{}upi\ifmmode~\acute{c}\else \'{c}\fi{}},\
  and\ \citenamefont {Branciard}}]{dourdent21}%
  \BibitemOpen
  \bibfield  {author} {\bibinfo {author} {\bibfnamefont {H.}~\bibnamefont
  {Dourdent}}, \bibinfo {author} {\bibfnamefont {A.~A.}\ \bibnamefont
  {Abbott}}, \bibinfo {author} {\bibfnamefont {N.}~\bibnamefont {Brunner}},
  \bibinfo {author} {\bibfnamefont {I.}~\bibnamefont {\ifmmode \check{S}\else
  \v{S}\fi{}upi\ifmmode~\acute{c}\else \'{c}\fi{}}},\ and\ \bibinfo {author}
  {\bibfnamefont {C.}~\bibnamefont {Branciard}},\ }\bibfield  {title} {\bibinfo
  {title} {Semi-device-independent certification of causal nonseparability with
  trusted quantum inputs},\ }\href
  {https://doi.org/10.1103/PhysRevLett.129.090402} {\bibfield  {journal}
  {\bibinfo  {journal} {Phys. Rev. Lett.}\ }\textbf {\bibinfo {volume} {129}},\
  \bibinfo {pages} {090402} (\bibinfo {year} {2022})},\ \Eprint
  {https://arxiv.org/abs/2107.10877} {arXiv:2107.10877 [quant-ph]} \BibitemShut
  {NoStop}%
\bibitem [{\citenamefont {Cavalcanti}\ \emph {et~al.}(2009)\citenamefont
  {Cavalcanti}, \citenamefont {Jones}, \citenamefont {Wiseman},\ and\
  \citenamefont {Reid}}]{cavalcanti09}%
  \BibitemOpen
  \bibfield  {author} {\bibinfo {author} {\bibfnamefont {E.~G.}\ \bibnamefont
  {Cavalcanti}}, \bibinfo {author} {\bibfnamefont {S.~J.}\ \bibnamefont
  {Jones}}, \bibinfo {author} {\bibfnamefont {H.~M.}\ \bibnamefont {Wiseman}},\
  and\ \bibinfo {author} {\bibfnamefont {M.~D.}\ \bibnamefont {Reid}},\
  }\bibfield  {title} {\bibinfo {title} {Experimental criteria for steering and
  the einstein-podolsky-rosen paradox},\ }\href
  {https://doi.org/10.1103/PhysRevA.80.032112} {\bibfield  {journal} {\bibinfo
  {journal} {Phys. Rev. A}\ }\textbf {\bibinfo {volume} {80}},\ \bibinfo
  {pages} {032112} (\bibinfo {year} {2009})},\ \Eprint
  {https://arxiv.org/abs/0907.1109} {arXiv:0907.1109 [quant-ph]} \BibitemShut
  {NoStop}%
\bibitem [{\citenamefont {Skrzypczyk}\ \emph {et~al.}(2014)\citenamefont
  {Skrzypczyk}, \citenamefont {Navascu\'es},\ and\ \citenamefont
  {Cavalcanti}}]{skrzypczyk14}%
  \BibitemOpen
  \bibfield  {author} {\bibinfo {author} {\bibfnamefont {P.}~\bibnamefont
  {Skrzypczyk}}, \bibinfo {author} {\bibfnamefont {M.}~\bibnamefont
  {Navascu\'es}},\ and\ \bibinfo {author} {\bibfnamefont {D.}~\bibnamefont
  {Cavalcanti}},\ }\bibfield  {title} {\bibinfo {title} {Quantifying
  einstein-podolsky-rosen steering},\ }\href
  {https://doi.org/10.1103/PhysRevLett.112.180404} {\bibfield  {journal}
  {\bibinfo  {journal} {Phys. Rev. Lett.}\ }\textbf {\bibinfo {volume} {112}},\
  \bibinfo {pages} {180404} (\bibinfo {year} {2014})},\ \Eprint
  {https://arxiv.org/abs/1311.4590} {arXiv:1311.4590 [quant-ph]} \BibitemShut
  {NoStop}%
\bibitem [{\citenamefont {de~Pillis}(1967)}]{depillis67}%
  \BibitemOpen
  \bibfield  {author} {\bibinfo {author} {\bibfnamefont {J.}~\bibnamefont
  {de~Pillis}},\ }\bibfield  {title} {\bibinfo {title} {Linear transformations
  which preserve hermitian and positive semidefinite operators},\ }\href
  {https://doi.org/10.2140/pjm.1967.23.129} {\bibfield  {journal} {\bibinfo
  {journal} {Pacific Journal of Mathematics}\ }\textbf {\bibinfo {volume}
  {23}},\ \bibinfo {pages} {129} (\bibinfo {year} {1967})}\BibitemShut
  {NoStop}%
\bibitem [{\citenamefont {Jamio\l{}kowski}(1972)}]{jamiolkowski72}%
  \BibitemOpen
  \bibfield  {author} {\bibinfo {author} {\bibfnamefont {A.}~\bibnamefont
  {Jamio\l{}kowski}},\ }\bibfield  {title} {\bibinfo {title} {Linear
  transformations which preserve trace and positive semidefiniteness of
  operators},\ }\href {https://doi.org/10.1016/0034-4877(72)90011-0} {\bibfield
   {journal} {\bibinfo  {journal} {Reports on Mathematical Physics}\ }\textbf
  {\bibinfo {volume} {3}},\ \bibinfo {pages} {275} (\bibinfo {year}
  {1972})}\BibitemShut {NoStop}%
\bibitem [{\citenamefont {Choi}(1975)}]{choi75}%
  \BibitemOpen
  \bibfield  {author} {\bibinfo {author} {\bibfnamefont {M.-D.}\ \bibnamefont
  {Choi}},\ }\bibfield  {title} {\bibinfo {title} {Completely positive linear
  maps on complex matrices},\ }\href
  {https://doi.org/10.1016/0024-3795(75)90075-0} {\bibfield  {journal}
  {\bibinfo  {journal} {Linear Algebra and its Applications}\ }\textbf
  {\bibinfo {volume} {10}},\ \bibinfo {pages} {285} (\bibinfo {year}
  {1975})}\BibitemShut {NoStop}%
\bibitem [{\citenamefont {Boyd}\ and\ \citenamefont
  {Vandenberghe}(2004)}]{boyd04}%
  \BibitemOpen
  \bibfield  {author} {\bibinfo {author} {\bibfnamefont {S.}~\bibnamefont
  {Boyd}}\ and\ \bibinfo {author} {\bibfnamefont {L.}~\bibnamefont
  {Vandenberghe}},\ }\href {http://stanford.edu/~boyd/cvxbook/} {\emph
  {\bibinfo {title} {Convex Optimization}}}\ (\bibinfo  {publisher} {Cambridge
  University Press},\ \bibinfo {year} {2004})\BibitemShut {NoStop}%
\bibitem [{\citenamefont {Bavaresco}(2022)}]{github_bavaresco}%
  \BibitemOpen
  \bibfield  {author} {\bibinfo {author} {\bibfnamefont {J.}~\bibnamefont
  {Bavaresco}},\ }\bibfield  {title} {\bibinfo {title} {{Code to accompany:
  ``Semi-device-independent certification of indefinite causal order in a
  photonic quantum switch''}},\ }\href
  {https://github.com/jessicabavaresco/experimental-SDI-causality} {\bibfield
  {journal} {\bibinfo  {journal}
  {https://github.com/jessicabavaresco/experimental-SDI-causality}\ } (\bibinfo
  {year} {2022})}\BibitemShut {NoStop}%
\bibitem [{\citenamefont {Rambo}\ \emph {et~al.}(2016)\citenamefont {Rambo},
  \citenamefont {Altepeter}, \citenamefont {Kumar},\ and\ \citenamefont
  {D'Ariano}}]{rambo2016functional}%
  \BibitemOpen
  \bibfield  {author} {\bibinfo {author} {\bibfnamefont {T.~M.}\ \bibnamefont
  {Rambo}}, \bibinfo {author} {\bibfnamefont {J.~B.}\ \bibnamefont
  {Altepeter}}, \bibinfo {author} {\bibfnamefont {P.}~\bibnamefont {Kumar}},\
  and\ \bibinfo {author} {\bibfnamefont {G.~M.}\ \bibnamefont {D'Ariano}},\
  }\bibfield  {title} {\bibinfo {title} {Functional quantum computing: An
  optical approach},\ }\href {https://doi.org/10.1103/PhysRevA.93.052321}
  {\bibfield  {journal} {\bibinfo  {journal} {Phys. Rev. A}\ }\textbf {\bibinfo
  {volume} {93}},\ \bibinfo {pages} {052321} (\bibinfo {year} {2016})},\
  \Eprint {https://arxiv.org/abs/1211.1257} {arXiv:1211.1257 [quant-ph]}
  \BibitemShut {NoStop}%
\bibitem [{\citenamefont {Wechs}\ \emph {et~al.}(2021)\citenamefont {Wechs},
  \citenamefont {Dourdent}, \citenamefont {Abbott},\ and\ \citenamefont
  {Branciard}}]{wechs21}%
  \BibitemOpen
  \bibfield  {author} {\bibinfo {author} {\bibfnamefont {J.}~\bibnamefont
  {Wechs}}, \bibinfo {author} {\bibfnamefont {H.}~\bibnamefont {Dourdent}},
  \bibinfo {author} {\bibfnamefont {A.~A.}\ \bibnamefont {Abbott}},\ and\
  \bibinfo {author} {\bibfnamefont {C.}~\bibnamefont {Branciard}},\ }\bibfield
  {title} {\bibinfo {title} {Quantum circuits with classical versus quantum
  control of causal order},\ }\href
  {https://doi.org/10.1103/PRXQuantum.2.030335} {\bibfield  {journal} {\bibinfo
   {journal} {PRX Quantum}\ }\textbf {\bibinfo {volume} {2}},\ \bibinfo {pages}
  {030335} (\bibinfo {year} {2021})},\ \Eprint
  {https://arxiv.org/abs/2101.08796} {arXiv:2101.08796 [quant-ph]} \BibitemShut
  {NoStop}%
\bibitem [{\citenamefont {Ara{\'{u}}jo}\ \emph
  {et~al.}(2014{\natexlab{b}})\citenamefont {Ara{\'{u}}jo}, \citenamefont
  {Costa},\ and\ \citenamefont {Brukner}}]{araujo2014computational}%
  \BibitemOpen
  \bibfield  {author} {\bibinfo {author} {\bibfnamefont {M.}~\bibnamefont
  {Ara{\'{u}}jo}}, \bibinfo {author} {\bibfnamefont {F.}~\bibnamefont
  {Costa}},\ and\ \bibinfo {author} {\bibfnamefont {{\v{C}}.}~\bibnamefont
  {Brukner}},\ }\bibfield  {title} {\bibinfo {title} {Computational advantage
  from quantum-controlled ordering of gates},\ }\href
  {https://doi.org/10.1103/PhysRevLett.113.250402} {\bibfield  {journal}
  {\bibinfo  {journal} {Phys. Rev. Lett.}\ }\textbf {\bibinfo {volume} {113}},\
  \bibinfo {pages} {250402} (\bibinfo {year} {2014}{\natexlab{b}})},\ \Eprint
  {https://arxiv.org/abs/1401.8127} {arXiv:1401.8127 [quant-ph]} \BibitemShut
  {NoStop}%
\end{thebibliography}


%



\onecolumngrid 
\appendix


\section*{APPENDIX}

The Appendix is organized as follows: In Sec.~\ref{app::SDPcausalmodels}, we define semi-device-independent causal models and inequalities, and their formulation as semidefinite programs. In Sec.~\ref{app::choioperators}, we describe the local operations and quantum switch process used in the calculation of the theoretical probability distributions. In Sec.~\ref{app::tomography}, we provide further details about the experimental characterization of Alice's operations.

\section{Semi-device-independent causal models and inequalities }\label{app::SDPcausalmodels}

Following Ref.~\cite{bavaresco19}, consider the scenario under the assumptions that the process, Bob's, and Charlie's operations are uncharacterized, and Alice's operations are fully characterized, a scenario referred to in Ref.~\cite{bavaresco19} as TUU (as in Trusted-Untrusted-Untrusted, referring to the partition Alice-Bob-Charlie, and using the word ``(un)trusted'' here as synonym to ``(un)characterized''). In this scenario, a causal model (also called a TUU-causal assemblage) is a set of operators $\{w^\text{causal}_{bc|yz}\}$, $w^\text{causal}_{bc|yz}\in\mathcal{L}(\mathcal{H}^{A_I}\otimes\mathcal{H}^{A_O})$ that recovers the statistics $\{p(abc|xyz)\}$ of an experiment that implemented the characterized instruments $\{\overline{A}_{a|x}\}$ according to
\begin{equation}
    p(abc|xyz) = \tr \left( \overline{A}_{a|x}\,w^\text{causal}_{bc|yz}\right).
\end{equation}
An experiment described by $\{p(abc|xyz)\}$ and $\{\overline{A}_{a|x}\}$ that can be recovered by a causal model as above, is one that can be simulated by a process that is causally separable, and therefore does not certify indefinite causal order~\cite{bavaresco19}.

A causal model $\{w^\text{causal}_{bc|yz}\}$, as per Ref.~\cite{bavaresco19}, is defined as below: 
\begin{equation}
    w^{\text{causal}}_{bc|yz} \coloneqq q w^{A\prec B\prec C}_{bc|yz} + (1-q) w^{B\prec A\prec C}_{bc|yz} \ \ \ \forall \ b,c,y,z,
\end{equation}
for some $0\leq q\leq 1$, where $\{w^{A\prec B\prec C}_{bc|yz}\}$ must satisfy
\begin{align}
    w^{A\prec B\prec C}_{bc|yz} &\geq 0 \ \ \  \forall \ b,c,y,z \\ 
    \tr \sum_{b,c} w^{A\prec B\prec C}_{bc|yz} &= d_{A_O} \ \ \  \forall \ y,z \\
    \sum_{c} w^{A\prec B\prec C}_{bc|yz} &= \sum_{c} w^{A\prec B\prec C}_{bc|yz'} \ \ \  \forall \ b,y,z,z' \\
    \sum_{b,c} w^{A\prec B\prec C}_{bc|yz} &= \sum_{b,c} w^{A\prec B\prec C}_{bc|y'z'} \ \ \  \forall \ y,y',z,z' \\
    \sum_{b,c} w^{A\prec B\prec C}_{bc|yz} &= \tr_{A_O} \sum_{b,c} w^{A\prec B\prec C}_{bc|yz} \otimes \frac{\id^{A_O}}{d_{A_O}} \ \ \  \forall \ y,z,
\end{align}
and $\{w^{B\prec A\prec C}_{bc|yz}\}$ must satisfy
\begin{align}
    w^{B\prec A\prec C}_{bc|yz} &\geq 0 \ \ \  \forall \ b,c,y,z \\
    \tr \sum_{b,c} w^{B\prec A\prec C}_{bc|yz} &= d_{A_O} \ \ \  \forall \ y,z \\
    \sum_{c} w^{B\prec A\prec C}_{bc|yz} &= \sum_{c} w^{B\prec A\prec C}_{bc|yz'} \ \ \  \forall \ b,y,z,z' \\
    \sum_{c} w^{B\prec A\prec C}_{bc|yz} &= \tr_{A_O} \sum_{c} w^{B\prec A\prec C}_{bc|yz} \otimes \frac{\id^{A_O}}{d_{A_O}} \ \ \  \forall \ b,y,z.
\end{align}
We refer to Ref.~\cite{bavaresco19} for details of the derivation of this causal model.

Given a set of probability distributions $\{p(abc|xyz)\}$ and a set of characterized instruments $\{\bar{A}_{a|x}\}$, the amount of randomness---which can be in this context interpreted as white noise---that can be mixed with $\{p(abc|xyz)\}$ such that it accepts a description by some causal model $\{w^\text{causal}_{bc|yz}\}$, is given by the solution of the following semidefinite program (SDP), which we call the primal problem:

\begin{flalign}\label{sdp::primal}
\begin{aligned}
    \textbf{given}\ \      &\{p(abc|xyz)\}, \{\bar{A}_{a|x}\} \\
    \textbf{maximize}\ \   & \eta \\ 
    \textbf{subject to}\ \ & \eta\,p(abc|xyz) + (1-\eta)\frac{1}{N_O}= \tr\left(\bar{A}_{a|x}\,w_{bc|yz}^\text{causal}\right) \ \ \ \forall \ a,b,c,x,y,z \\
                           &\{w_{bc|yz}^\text{causal}\}\ \text{is a causal model},
\end{aligned}&&
\end{flalign}
where $N_O$ is the total number of outcomes of the experiment and the optimization is taken over the variables $\eta$ and $\{w^\text{causal}_{bc|yz}\}$. If the solution of this SDP is $\max\{\eta\}\geq1$, then a causal model exists and indefinite causal order cannot be certified in the experiment described by $\{p(abc|xyz)\}$ and $\{\bar{A}_{a|x}\}$. Alternatively, if $\max\{\eta\}<1$, then one certifies that the experiment described by $\{p(abc|xyz)\}$ and $\{\bar{A}_{a|x}\}$ does not accept a causal model and therefore demonstrated indefinite causal order.

The dual problem associated to the above SDP is given by
\begin{flalign}\label{sdp::dual}
\begin{aligned}
    \textbf{given}\ \      &\{p(abc|xyz)\}, \{\bar{A}_{a|x}\} \\
    \textbf{minimize}\ \   & 1 + \sum_{\substack{abc\\xyz}}\,\alpha^{abc}_{xyz}\,p(abc|xyz)\\ 
    \textbf{subject to}\ \ & \sum_{ax}\alpha^{abc}_{xyz}\,\bar{A}_{a|x} - \sigma^{A\prec B\prec C}_{b|yz}\geq0 \ \ \ \forall \ b,c,y,z\\
                           & \sum_{ax}\alpha^{abc}_{xyz}\,\bar{A}_{a|x} - \sigma^{B\prec A\prec C}_{b|yz}\geq0 \ \ \ \forall \ b,c,y,z\\
                           & \frac{1}{N_O} \sum_{\substack{abc\\xyz}}\, \alpha^{abc}_{xyz} = 1 + \sum_{\substack{abc\\xyz}}\,\alpha^{abc}_{xyz}\,p(abc|xyz),
\end{aligned}&&
\end{flalign}
where the optimization is taken over the variables $\{\alpha^{abc}_{xyz}\}$, which is a set of real coefficients, and the variables $\{\sigma^{A\prec B\prec C}_{b|yz}\}$ and $\{\sigma^{B\prec A\prec C}_{b|yz}\}$, which are sets of operators given by
\begin{align}
    \sigma^{A\prec B\prec C}_{b|yz} &= h^{A\prec B\prec C}_{yz}\id + K^{A\prec B\prec C}_{b|yz} + G^{A\prec B\prec C}_{yz} - \tr_{A_O}G^{A\prec B\prec C}_{yz}\otimes\frac{\id}{d_{A_O}} + R^{A\prec B\prec C}_{yz} \\
    \sigma^{B\prec A\prec C}_{b|yz} &= h^{B\prec A\prec C}_{yz}\id + K^{B\prec A\prec C}_{b|yz} + J^{B\prec A\prec C}_{b|yz} - \tr_{A_O}J^{B\prec A\prec C}_{b|yz}\otimes\frac{\id}{d_{A_O}},
\end{align}
for all $b,y,z$, and where $\{h^{A\prec B\prec C}_{yz}\}$, $\{h^{B\prec A\prec C}_{yz}\}$, $\{K^{A\prec B\prec C}_{yz}\}$, $\{K^{B\prec A\prec C}_{yz}\}$, and $\{R^{A\prec B\prec C}_{yz}\}$ must satisfy
\begin{align}
    \sum_{yz}h^{A\prec B\prec C}_{yz}=0&, \hspace*{1cm} \sum_{yz}h^{B\prec A\prec C}_{yz}=0 \\ 
    \sum_{z}K^{A\prec B\prec C}_{b|yz}=0&, \hspace*{1cm} \sum_{z}K^{B\prec A\prec C}_{b|yz}=0 \ \ \ \forall\ b,y \\
    \sum_{yz}R^{A\prec B\prec C}_{yz}=0&,
\end{align}
and the other variables can be any complex hermitian matrices.

Since primal and dual SDPs satisfy the condition of strong duality~\cite{boyd04}, it is known that their solutions coincide. Hence, $\max\{\eta\}=\min\{1 + \sum_{\substack{abc\\xyz}}\,\alpha^{abc}_{xyz}\,p(abc|xyz)\}\geq1$ implies the existence of a causal model, while $\max\{\eta\}=\min\{1 + \sum_{\substack{abc\\xyz}}\,\alpha^{abc}_{xyz}\,p(abc|xyz)\}<1$ implies indefinite causal order, leading to the inequality
\begin{equation}
    S \coloneqq \sum_{\substack{abc\\xyz}}\,\alpha^{abc}_{xyz}\,p(abc|xyz) \geq 0,
\end{equation}
derived under semi-device-independent assumptions, which is satisfied if and only if the experiment described by $\{p(abc|xyz)\}$ and $\{\bar{A}_{a|x}\}$ can be explained by a causal model. The coefficients $\{\alpha^{abc}_{xyz}\}$ of the above inequality are obtained from the solution of the dual problem, and they will depend on both $\{p(abc|xyz)\}$ and $\{\bar{A}_{a|x}\}$.


\section{The quantum switch process and local operations}\label{app::choioperators}

In this section, we specify exactly what are the local operations that Alice is assumed to experimentally implement. In Sec.~\ref{app::tomography}, the reader will find details on the process tomography performed on Alice's local operations to ensure that the experimentally implemented operations indeed correspond to the theoretically assumed ones, described below.

Furthermore, in this section we also compute a theoretical prediction for the sets of probability distributions that can be measured in our experiment. For this purpose, we use a description of the quantum switch process and a choice of local operations for Bob and Charlie described below. Crucially, these operations are only used to calculate a theoretical prediction for the probability distributions and are not assumed in the analysis of the experimental data or evaluation of the inequality violation.

As described in the main text, this theoretical prediction of the probability distributions allows us to derive, using the SDP in Sec.~\ref{app::SDPcausalmodels}, the coefficients of a tailored semi-device-independent causal inequality that is expected to capture the indefinite-causal-order properties demonstrated in our experimental setup.

Within the process matrix formalism~\cite{araujo15}, the quantum switch can be described as an operator $W^\text{switch}\in\mathcal{L}(\mathcal{H}^{A_I}\otimes\mathcal{H}^{A_O}\otimes\mathcal{H}^{B_I}\otimes\mathcal{H}^{B_O}\otimes\mathcal{H}^{C_t}\otimes\mathcal{H}^{C_c})$ that acts on the linear spaces of Alice's and Bob's input ($A_I,B_I$) and output ($A_O,B_O$) systems, and of Charlie's future target ($C_t$) and control ($C_c$) systems. A quantum switch that has a control system in the initial state $\ket{+}=(\ket{0}+\ket{1})/\sqrt{2}$ and a target system in the initial state $\ket{0}$ is given be the operator 
\begin{equation}\label{appeq::switch}
W^\text{switch}=\ketbra{w}{w}, 
\end{equation}
where
\begin{equation}
    \ket{w} = \frac{1}{\sqrt{2}}(\ket{0}^{A_I}\ket{\id}^{A_OB_I}\ket{\id}^{B_OC_t}\ket{0}^{C_c} + \ket{0}^{B_I}\ket{\id}^{B_OA_I}\ket{\id}^{A_OC_t}\ket{1}^{C_c}),
\end{equation}
and $\ket{\id}^{IO}=\sum_i\ket{i}^I\ket{i}^O$.

The local instruments performed by Alice and Bob are given by their corresponding Choi operators below. The measurements performed by Charlie are also described below.
The instruments of Alice are given by a set of operators $\{A_{a|x}\}$, $A_{a|x}\in\mathcal{L}(\mathcal{H}^{A_I}\otimes\mathcal{H}^{A_O})$ with $a,x\in\{0,1\}$, such that
\begin{align}
A_{0|0} &= \ketbra{0}{0}^{A_I}\otimes\ketbra{0}{0}^{A_O} \label{eq::alice00}\\
A_{1|0} &= \ketbra{1}{1}^{A_I}\otimes\ketbra{1}{1}^{A_O} \label{eq::alice10}\\
A_{0|1} &= \ketbra{+}{+}^{A_I}\otimes\ketbra{+}{+}^{A_O} \label{eq::alice01}\\
A_{1|1} &= \ketbra{-}{-}^{A_I}\otimes\ketbra{-}{-}^{A_O}, \label{eq::alice11}
\end{align}
where $\ket{\pm}\coloneqq(\ket{0}\pm\ket{1})/\sqrt{2}$. Essentially, these instruments correspond to Alice first measuring the target qubit on either the Pauli $Z$ or $X$ basis and then re-preparing the eigenstate corresponding to her measurement outcome.

Bob's instruments are identical to Alice's, i.e., $\{B_{b|y}\}$, $B_{b|y}\in\mathcal{L}(\mathcal{H}^{B_I}\otimes\mathcal{H}^{B_O})$ with $b,y\in\{0,1\}$, that is,
\begin{align}
B_{0|0} &= A_{0|0} \label{appeq::bs} \\
B_{1|0} &= A_{1|0}  \\
B_{0|1} &= A_{0|1}  \\
B_{1|1} &= A_{1|1}.\label{appeq::be}
\end{align}

Finally, the measurements of Charlie are defined as $\{M_{c|z}\}$, $M_{c|z}\in\mathcal{L}(\mathcal{H}^{C_t}\otimes\mathcal{H}^{C_c})$ with $z\in\{0,1\}$ and $c\in\{0,1,2,3\}$, given by
\begin{align}
M_{0|0} &= \ketbra{0}{0}^{C_t}\otimes\ketbra{+}{+}^{C_c}, \label{appeq::cs} \\
M_{1|0} &= \ketbra{1}{1}^{C_t}\otimes\ketbra{+}{+}^{C_c}, \\
M_{2|0} &= \ketbra{0}{0}^{C_t}\otimes\ketbra{-}{-}^{C_c}, \\
M_{3|0} &= \ketbra{1}{1}^{C_t}\otimes\ketbra{-}{-}^{C_c}, \\ 
\nonumber \\
M_{0|1} &= \ketbra{+}{+}^{C_t}\otimes\ketbra{+}{+}^{C_c}, \\
M_{1|1} &= \ketbra{+}{+}^{C_t}\otimes\ketbra{-}{-}^{C_c}, \\
M_{2|1} &= \ketbra{-}{-}^{C_t}\otimes\ketbra{+}{+}^{C_c}, \\
M_{3|1} &= \ketbra{-}{-}^{C_t}\otimes\ketbra{-}{-}^{C_c}. \label{appeq::ce}
\end{align}
Charlie's first measurement corresponds to performing a projective measurement on the Pauli $Z$ basis on the target qubit and on the Pauli $X$ basis on the control qubit, while his second measurement corresponds to performing a projective measurement on the Pauli $X$ basis on both target and control qubits.

The theoretical probability distributions $\{p_\text{theory}(abc|xyz)\}$ are then calculated from $W^\text{switch}$, $\{A_{a|x}\}$, $\{B_{b|y}\}$, and $\{M_{c|z}\}$ given above, according to
\begin{equation}
    p_\text{theory}(abc|xyz) = \tr\left[\left(A_{a|x}\otimes B_{b|y}\otimes M_{c|z}\right)\,W^\text{switch}\right], \ \ \ \forall \ a,b,c,x,y,z.
\end{equation}

By evaluating SDP~\eqref{sdp::dual} with the probability distributions $\{p_\text{theory}(abc|xyz)\}$ and the set of local instruments for Alice $\{A_{a|x}\}$ described above as input, one obtains the coefficients $\{\alpha^{abc}_{xyz}\}$ of the semi-device-independent causal inequality tested in our experiment. We also obtain a theoretical value for the inequality score of $S_\text{theory}=-0.0794$.

We would also like to remark that, although the operations above were the ones used in the computation of our theoretical prediction of the sets of probability distributions, they are not the only possible set of local operations that can lead to a semi-device-independent certification of indefinite causal order. Take for example the set of probability distributions that can be computed from the quantum switch process $W^\text{switch}$ in Eq.~\eqref{appeq::switch}, the same set of instruments for Bob as in Eqs.~\eqref{appeq::bs}-\eqref{appeq::be}, the same set of measurements for Charlie as in Eqs.~\eqref{appeq::cs}--\eqref{appeq::ce}, and for Alice, three unitary (single-outcome) operations that act on the target system according to $A_{1}(\rho) = \sigma_X\,\rho\,\sigma_X$, $A_{2}(\rho) = \sigma_Y\,\rho\,\sigma_Y$, and $A_{3}(\rho) = \ketbra{0}{0}$, where $\sigma_X$ and $\sigma_Y$ are Pauli operators, and $A_{3}$ is a trace-and-replace map that discards the input state and deterministically prepares the output state $\ketbra{0}{0}$. By evaluating SDP~\eqref{sdp::dual} with this set of probability distributions and this set of local operations for Alice, it can be checked that the corresponding semi-device-independent inequality would be violated by a value of $S_\text{theory}=-0.0400$.


\section{Characterization of Alice's instruments}\label{app::tomography}

The theoretical model of our work is described with the process matrix formalism. The instrument elements of all local parties can be described by their corresponding Choi states. The main merit in our work is to assumed that only the instruments of Alice are characterized, while those of Bob and Charlie are treated device-independently. To characterize the operations of Alice's side and check that she actually performs the desired instruments---justifying the device-dependent assumption of her operations---we experimentally performed process tomography.

\begin{figure}
\begin{center}
    \includegraphics[width=\columnwidth]{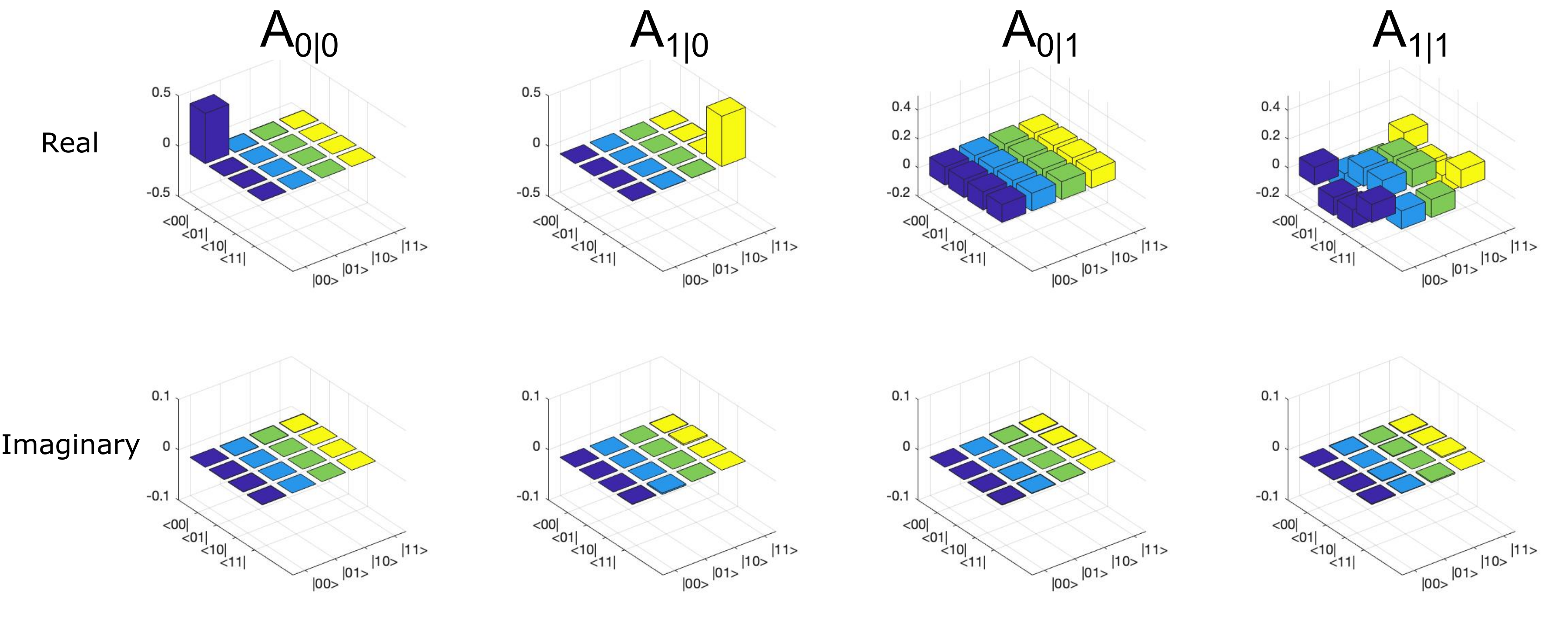}
    \caption{Choi states of the characterized party's instruments. The basis of the Choi states is $\{\ket{i}^{A_I}\otimes\ket{j}^{A_O}\}_{ij}$, $i,j\in\{0,1\}$. Each column describes an instrument element corresponding to a measure-and-reprepare operation with outcome $0$ in the $\sigma_z$ basis, outcome $1$ in the $\sigma_z$ basis, outcome $0$ in the $\sigma_X$ basis, and outcome $1$ in the $\sigma_X$ basis, in this order. The real part of the Choi state is presented in the first row and the imaginary part in the second row.}
    \label{fig:Choi state}
\end{center}
\end{figure}

Figure~\ref{fig:Choi state} illustrates the Choi states of Alice's instrument elements. The four Choi states correspond to each of Alice's possible measurement basis in $\sigma_x$ and $\sigma_z$ direction with possible outcomes $a\in\{0,1\}$. The Choi states of each instrument element exhibited fidelity of $\{0.9989,0.9991,0.9991,0.9990\}$, with respect to ideal ones $\{A_{0|0},A_{0|1},A_{1|0},A_{1|1}\}$ given by Eqs.~\eqref{eq::alice00}--\eqref{eq::alice11}, respectively. The errorbars are exempted since we collected a sufficiently high number of counts to make them negligible. The high fidelities we measured justify the validity of our characterization assumption over Alice's operations.

\end{document}